\documentclass[english,prl,reprint,showpacs]{revtex4-1}
\usepackage[T1]{fontenc}
\usepackage[latin9]{inputenc}
\usepackage{babel}
\usepackage{amsmath}
\usepackage{wasysym}
\usepackage{graphicx}
\usepackage{esint}
\usepackage[unicode=true,pdfusetitle,
 bookmarks=true,bookmarksnumbered=false,bookmarksopen=false,
 breaklinks=false,pdfborder={0 0 0},backref=false,colorlinks=false]
 {hyperref}



\begin{document}
\begin{abstract}
We experimentally investigate and utilize electrothermal feedback
in a microwave nanobolometer based on a normal-metal ($\textnormal{Au}_{x}\textnormal{Pd}_{1-x}$)
nanowire with proximity-induced superconductivity.  The feedback
couples the temperature and the electrical degrees of freedom in the
 nanowire, which both absorbs the incoming microwave radiation, and
transduces the temperature change into a radio-frequency electrical
signal.  We tune the feedback in situ and access both positive and
negative feedback regimes with rich nonlinear dynamics. In particular,
strong positive feedback leads to the emergence of two metastable
electron temperature states in the millikelvin range. We use these
states for efficient threshold detection of coherent $8.4\mbox{ GHz}$
microwave pulses containing approximately 200 photons on average,
corresponding to $1.1\times10^{-21}\textnormal{ J}\approx7.0\textnormal{ meV}$
of energy. 
\end{abstract}

\pacs{07.57.Kp, 74.78.Na, 74.45.+c, 85.25.Cp }

\title{Detection of zeptojoule microwave pulses using electrothermal feedback
in proximity-induced Josephson junctions}

\author{J. Govenius}

\email{joonas.govenius@aalto.fi}

\author{R. E. Lake}

\author{K. Y. Tan}

\author{M. Möttönen}

\affiliation{QCD Labs, COMP Centre of Excellence, Department of Applied Physics,
Aalto University, P.O. Box 13500, FIN-00076 Aalto, Finland}

\maketitle
Superconducting qubits coupled to microwave transmission lines have
developed into a versatile platform for solid-state quantum optics
experiments \cite{Blais04,Wallraff2004Strong}, as well as a promising
candidate for quantum computing \cite{Devoret2013Superconducting,Kelly2015State}.
 However, compared to optical photodetectors \cite{Lita2008Counting,Marsili2013Detecting,Eisaman2011Invited},
detectors for itinerant single-photon microwave pulses are still in
their infancy. This prevents microwave implementations of optical
protocols that require feedback conditioned on single-photon detection
events. For example, linear optical quantum computing  with single-photon
pulses calls for such feedback \cite{Kok2007Linear}. Photodetection
and feedback can also act as a quantum eraser \cite{Hillery1983State}
of the phase information available in a coherent signal, as we recently
discussed in Ref. \cite{Govenius2015Parity}. Note that, given sufficient
averaging, linear amplifiers can substitute for photodetectors in
ensemble-averaged experiments \cite{daSilva2010Schemes,Bozyigit2010Antibunching},
but the uncertainty principle fundamentally limits the success probability
in single-shot experiments.  

 We focus on thermal photodetectors, i.e., detectors that measure
the temperature rise caused by absorbed photons. Thermal detectors
have been developed for increasingly long wavelengths in the context
of THz astronomy \cite{Karasik2011Nanobolometers}, the record being
the detection of single $8\textnormal{ \ensuremath{\mu}m}$ photons
\cite{Karasik2012Energyresolved}. In the context of quantum thermodynamics
\cite{Pekola2015Towards}, thermal detectors have recently been proposed
\cite{Pekola2013Calorimetric} and developed \cite{Schmidt2003Nanoscale,Schmidt2005Superconductorinsulatornormal,Gasparinetti2015Fast,Saira2016Dispersive}
as monitorable heat baths. 

The other main approach to detecting itinerant microwave photons is
to use a qubit that is excited by an incoming photon and then measured
\cite{Romero2009Microwave,Chen2011Microwave,Peropadre2011Approaching,Fan2013Breakdown,Hoi2013Giant,Sathyamoorthy2014Quantum,Fan2014Nonabsorbing,Koshino2015Theory,Inomata2016Single,Narla2016Robust}.
Very recently, Ref.~\cite{Inomata2016Single} reported reaching an
efficiency of 0.66 and a bandwidth of roughly 20 MHz using such an
approach. Use of qubit-based single-photon transistors as photodetectors
has also been proposed \cite{Neumeier2013SinglePhoton,Manzoni2014Singlephoton}.
If the pulse is carefully shaped, it is also possible to efficiently
absorb a photon into a resonator~\cite{Cirac1997Quantum,Yin2013Catch,Srinivasan2014Timereversal,Wenner2014Catching,Pechal2014MicrowaveControlled}.
There it could be detected with established techniques for intra-resonator
photon counting \cite{Gleyzes2007Quantum,Sun2014Tracking}.

\begin{figure}[t]
\begin{centering}
\includegraphics{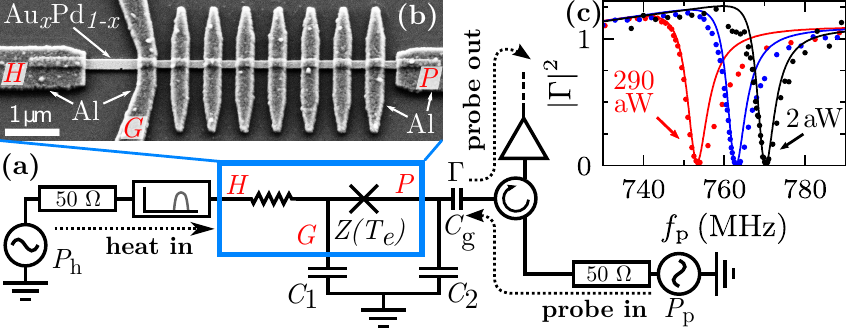}
\par\end{centering}

\caption{(Color online) (a) Simplified diagram of the detector, including (b)
a micrograph of the SNS junctions formed by a $\textnormal{Au}_{x}\textnormal{Pd}_{1-x}$
nanowire contacted by $\mbox{Al}$ islands and leads ($H$, $P$,
and $G$). Here, $Z^{-1}$ is an admittance, $T_{e}$ is the temperature
of the electrons in the nanowire, and $\Gamma$ is the probe signal
reflection coefficient. The micrograph is from a device nominally
identical to the measured one.  (c) Reflected fraction of probe power
versus probe frequency $f_{\textnormal{p}}$ for steady-state heating
power $P_{\textnormal{h}}$ of $1.9$, $66$, and $290\mbox{ aW}$.
They are measured at low probe power $P_{\textnormal{p}}\ll P_{\textnormal{h}}$.
The solid curves are fits to the circuit model with a small phenomenological
correction term \cite{Note1}. The heater input is bandpass filtered
($8.41\pm0.02\textnormal{ GHz})$.\label{fig:setup} }
\end{figure}

The main advantage of thermal detectors is that they typically present
a suitable real input impedance for absorbing photons efficiently
over a wide bandwidth and a large dynamic range, in contrast to qubit-based
detectors. However, a central problem in the thermal approach is the
small temperature rise caused by individual microwave photons. The
resulting transient temperature spike is easily overwhelmed by noise
added in the readout stage. One potential solution is to use a bistable
system as a threshold detector that maps a weak transient input pulse
to a long-lived metastable state of the detector. This is conceptually
similar to, e.g., early experiments on superconducting qubits that
used a current-biased superconducting quantum interference device
(SQUID) \cite{Chiorescu2003Coherent}. Conditioned on the initial
qubit state, the SQUID either remained in the superconducting state
or switched to a long-lived non-zero voltage state.

In this Letter, we show that an electrothermal bistability emerges
in the microwave nanobolometer we introduced in Ref.~\cite{Govenius2014Microwave}
and that it enables high-fidelity threshold detection of $8.4\mbox{ GHz}$
microwave pulses containing only $200\times h\times8.4\,\textnormal{GHz}\approx1.1\textnormal{ zJ}$
of energy. This threshold is more than an order of magnitude improvement
over previous thermal detector results \cite{Santavicca2010Energy,Karasik2012Energyresolved}.
The bistability in our detector arises from the fact that the amount
of power absorbed from the electrical probe signal used for readout
depends on the measured electron temperature itself. Previously, such
electrothermal feedback and the associated bifurcation has been studied
in the context of kinetic inductance detectors \cite{deVisser2010Readoutpower,Thompson2013Dynamical,Lindeman2014Resonatorbolometer,Thomas2015Electrothermal}.
Analogous thermal effects in optics are also known \cite{Braginsky1989Qualityfactor,Fomin2005Nonstationary}.
The main difference to our device is the relative strength of the
electrothermal effect, which in our case leads to strongly nonlinear
behavior at attowatt probe powers. Electrothermal feedback is also
commonly used in transition edge sensors \cite{Karasik2011Nanobolometers},
but typically the feedback is chosen to be negative because that suppresses
Johnson noise and leads to fast self-resetting behavior \cite{Irwin1995Application}.

The central component of our detector (Fig.~\ref{fig:setup}) is
a metallic $\textnormal{Au}_{x}\textnormal{Pd}_{1-x}$ nanowire ($x\approx0.6$)
contacted by three $\textnormal{Al}$ leads ($H$, $P$, and $G$)
and seven $\textnormal{Al}$ islands that are superconducting at millikelvin
temperatures %
\footnote{See the Supplemental Material for additional single-shot histograms
and for details of the experimental setup, the circuit model, the
numerical model, and the measurements of $\tilde{G}$, $\tau$, and
$C_{e}$.%
}. The longest superconductor--normal-metal--superconductor (SNS) junction
($H$--$G$) provides a resistive load ($36\textnormal{ \ensuremath{\Omega}}$)
for the radiation to be detected \cite{Nahum1993Ultrasensitivehotelectron},
while the shorter junctions ($P$--$G$) function as a proximity Josephson
sensor \cite{Giazotto2008Ultrasensitive,Voutilainen2010Physics}.
That is, the shorter junctions provide a temperature-dependent inductance
in an effective LC resonator used for readout. Because the inductance
increases with electron temperature $T_{e}$ in the nanowire, the
resonance frequency  shifts down as the heating power $P_{\textnormal{h}}$
increases. Therefore the detector transduces changes in $P_{\textnormal{h}}$
into changes in the reflection coefficient $\Gamma$ {[}Fig.~\ref{fig:setup}(c){]}.
 For simplicity, we limit the bandwidth of the heater line using
a Lorentzian bandpass filter, but replacing it with a wider band filter
should be straightforward.

We first characterize the detector by measuring the $P_{\textnormal{h}}$-dependence
of the admittance $Z^{-1}$ between $P$ and $G$. To do so, we fit
the measured $\Gamma$ to a circuit model in which we parametrize
$Z^{-1}$ as $R^{-1}+(i\omega L)^{-1}$, where $\omega/2\pi=f_{\textnormal{p}}$
is the probe frequency. The circuit model shown in Fig.~\ref{fig:setup}(a)
predicts $\Gamma=(Z_{\textnormal{L}}-Z_{0})/(Z_{\textnormal{L}}+Z_{0})$,
where 
\[
Z_{\textnormal{L}}=(i\omega C_{\textnormal{g}})^{-1}+\{i\omega C_{2}+[(i\omega C_{1})^{-1}+Z(T_{e})]^{-1}\}^{-1},
\]
$Z_{0}=50\textnormal{ \ensuremath{\Omega}}$, $C_{1}\approx87\textnormal{ pF}$,
$C_{2}\approx70\textnormal{ pF}$, and $C_{\textnormal{g}}\approx1.5\textnormal{ pF}$.
This fits reasonably well to the linear response data shown in Fig.~\ref{fig:setup}(c).
However, in order to reproduce the asymmetry in the measured lineshape,
we add a small frequency-dependent correction to the model \cite{Note1}.
Here, linear response refers to the use of a probe power $P_{\textnormal{p}}$
low enough to ignore both the electrical and electrothermal nonlinearities,
i.e., the nonlinearity of the Josephson inductance as well as the
variation of $T_{e}$ as a function of the absorbed probe power $(1-\left|\Gamma\right|^{2})P_{\textnormal{p}}$.
We note that the uncertainty in $P_{\textnormal{h}}$ is roughly $1\textnormal{ dB}$
\cite{Note1}, and that dissipation in the capacitors is negligible.

\begin{figure}[!t]
\begin{centering}
\includegraphics{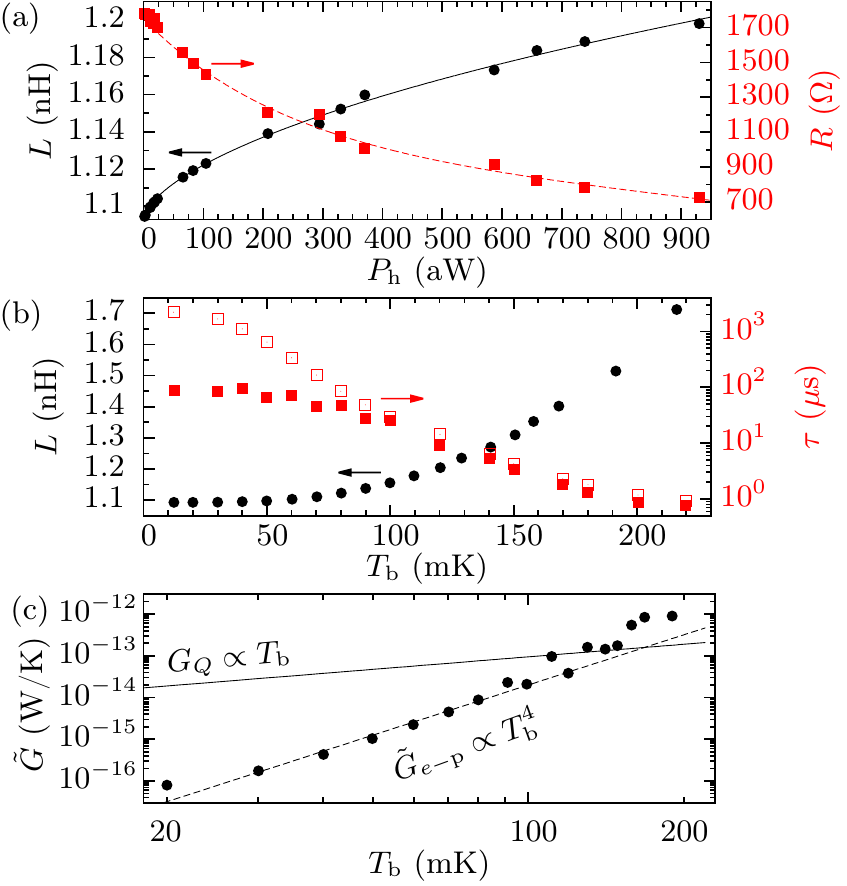}
\par\end{centering}

\caption{(Color online) Linear ($P_{\textnormal{p}}\ll P_{\textnormal{h}}$
) response. (a) The effective inductance (circles) and resistance
(squares) of the short SNS junctions as functions of external steady-state
heating power $P_{\textnormal{h}}$. The bath temperature $T_{\textnormal{b}}$
is $12\textnormal{ mK}$.  The curves are phenomenological fits that
allow mapping a measured reflection coefficient into an equivalent
$P_{\textnormal{h}}$. (b) The effective inductance (circles) and
thermal relaxation time after a short (filled squares) or long (open
squares) heating pulse. (c) Measured differential thermal conductance
$\tilde{G}$, the expected electron-phonon contribution $\tilde{G}_{e-\textnormal{p}}$
(dashed line), and the quantum of thermal conductance $G_{Q}$ (solid
line).  \label{fig:delta-calibration} }
\end{figure}

Figure~\ref{fig:delta-calibration}(a) shows the extracted linear
response $L$ and $R$ for heating powers up to a femtowatt. Figure~\ref{fig:delta-calibration}
also shows the bath temperature dependence of $L$, the thermal relaxation
time $\tau$, and the differential thermal conductance $\tilde{G}=-\partial_{T_{\textnormal{b}}}P_{e-\textnormal{b}}\left(T_{e},T_{\textnormal{b}}\right)$
\cite{Note1}. Here, $P_{e-\textnormal{b}}\left(T_{e},T_{\textnormal{b}}\right)$
is the heat flow between the electrons in the nanowire and the cryostat
phonon bath at temperature $T_{\textnormal{b}}$. The measured $\tilde{G}$
is in rough agreement with the prediction for electron--phonon limited
thermalization $\tilde{G}_{e-\textnormal{p}}=5\Sigma V_{0}T_{\textnormal{b}}^{4}$,
where $\Sigma\approx3\times10^{9}\textnormal{ W/m}^{3}\textnormal{K}^{5}$
is a material parameter \cite{Timofeev2009Electronic} and $V_{0}\approx\left(240\textnormal{ nm}\right)^{3}$
is the volume of the part of the nanowire not covered by Al. We can
use these results to estimate $C_{e}$ above $100\textnormal{ mK}$,
where $T_{e}\approx T_{\textnormal{b}}$ and $C_{e}\approx\tau\tilde{G}\approx\gamma V_{0}T_{\textnormal{b}}$,
with $\gamma V_{0}=8\textnormal{ aJ/K}^{2}$ \cite{Note1}.

Below $100\textnormal{ mK}$,  the relaxation toward the stationary
state is faster after a short ($1\textnormal{ \ensuremath{\mu}s}$)
heating pulse than after a long ($\gg\tau$) heating pulse \cite{Note1}.
Therefore, the simplest thermal model of a single heat capacity $C_{e}$
coupled directly to the bath is not accurate below $100\textnormal{ mK}$.
Instead, the second time scale can be phenomenologically explained
by an additional heat capacity $C^{\prime}\gg C_{e}$ coupled strongly
to $C_{e}$ but weakly to the bath, as compared to $\tilde{G}$. 
Since $\tilde{G}$ falls far below the quantum of thermal conductance
$G_{Q}=\pi^{2}k_{\textnormal{B}}^{2}T_{\textnormal{b}}/3h$ \cite{Pendry1983Quantum}
at low temperatures {[}Fig.~\ref{fig:delta-calibration}(c){]}, even
weak residual electromagnetic coupling \cite{Schmidt2004PhotonMediated,Meschke2006Singlemode,Partanen2016Quantumlimited}
between $C_{e}$ and $C^{\prime}$ would suffice. However, we cannot
uniquely determine the microscopic origin of $C^{\prime}$ or the
coupling mechanism from the data. Also note that a similar second
time scale was observed in Ref.~\cite{Gasparinetti2015Fast}.

At high probe powers, the linear-response behavior studied above may
be drastically modified by the absorbed probe power. Below we focus
on the stationary $T_{e}$ solutions, so we choose to neglect the
transient heat flows to $C^{\prime}$ that give rise to the shorter
time scale in Fig.~\ref{fig:delta-calibration}(b). Similarly, we
neglect the contribution of electrical transients to $\Gamma$, as
they decay even faster (in $\apprle100\,\mbox{ns}$). Under these
approximations, $T_{e}$ is the only dynamic variable and evolves
according to

\begin{align}
C_{e}\left(T_{e}\right)\dot{T}_{e} & =-P_{e-\textnormal{b}}\left(T_{e},T_{\textnormal{b}}\right)+P_{x}+P_{\textnormal{h}}\label{eq:dTedt}\\
 & \qquad+(1-\left|\Gamma\left(T_{e},\omega\right)\right|^{2})P_{\textnormal{p}},\nonumber 
\end{align}
where $P_{x}$ accounts for the average heat load from uncontrolled
sources.

Determining $T_{e}$ from Eq.~(\ref{eq:dTedt}) and the measured
$\Gamma$ would require additional assumptions about $P_{e-\textnormal{b}}$
and $P_{x}$, as they are not directly measurable. However, we avoid
making such assumptions by instead analyzing the increase in the heat
flow from the electrons to the thermal bath, as compared to the case
$P_{\textnormal{h}}=P_{\textnormal{p}}=0$. That is, instead of $T_{e}$,
we analyze

\begin{equation}
\Delta(T_{e})=P_{e-\textnormal{b}}(T_{e},T_{\textnormal{b}})-P_{x},\label{eq:Delta}
\end{equation}
which is monotonic in $T_{e}$. Given this definition, we can rewrite
Eq.~(\ref{eq:dTedt}) as
\begin{equation}
\tau\left(\Delta\right)\dot{\Delta}=-\Delta+P_{\textnormal{h}}+(1-\left|\Gamma\left(\Delta,\omega\right)\right|^{2})P_{\textnormal{p}},\label{eq:dDeltadt}
\end{equation}
where $\tau\left(\Delta\right)=C\left(\Delta\right)\slash\partial_{T_{e}}P_{e-\textnormal{b}}\textnormal{\textbf{(}}T_{e}(\Delta),T_{\textnormal{b}}\textnormal{\textbf{)}}$.
In contrast to the unknown parameters in Eq.~(\ref{eq:dTedt}), $\Gamma\left(\Delta,\omega\right)$
and $\tau\left(\Delta\right)$ are directly measurable in linear response.
Specifically, we can determine $\Gamma\left(\Delta,\omega\right)$
from the data in Fig.~\ref{fig:delta-calibration}(a) since $\Delta=P_{\textnormal{h}}$
when $P_{\textnormal{p}},\dot{\Delta}\rightarrow0$. By inverting
$\Gamma\left(\Delta,\omega\right)$, we can then extract $\Delta$
from the measured $\Gamma$. Also note that, since all parameters
in Eq.~(\ref{eq:dDeltadt}) are determined in linear response, no
free parameters remain in the theoretical predictions for the nonlinear
case discussed below.

\begin{figure}[!t]
\begin{centering}
\includegraphics{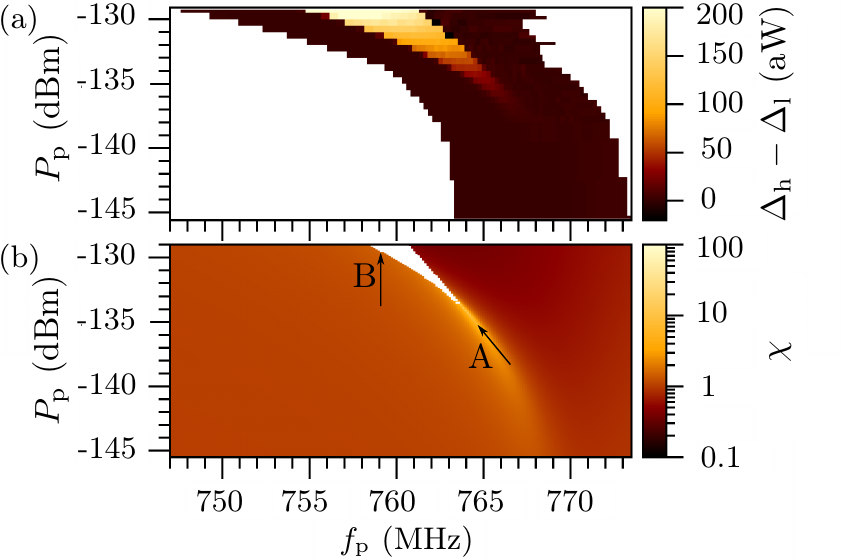}
\par\end{centering}

\caption{(Color online) (a) Bistable parameter regime, as indicated by a non-zero
difference $\Delta_{\textnormal{h}}-\Delta_{\textnormal{l}}$ in the
power absorbed from the probe signal in high and low-temperature stationary
states.   (b) Numerically simulated values of the dimensionless
susceptibility to external heating $\chi$ in the single-valued regime.
The bistable regime is indicated in white. \label{fig:phase-diag} }
\end{figure}

The emergence of bistability is the most dramatic consequence of increasing
the probe power. Experimentally, we map out the bistable parameter
regime by measuring the difference $\Delta_{\textnormal{h}}-\Delta_{\textnormal{l}}$
as a function of $f_{\textnormal{p}}$ and $P_{\textnormal{p}}$ {[}Fig.~\ref{fig:phase-diag}(a){]}.
Here, $\Delta_{\textnormal{h}}$ $\left(\Delta_{\textnormal{l}}\right)$
corresponds to the ensemble-averaged $\Delta$ measured $5\textnormal{ ms}$
after preparing the system in a high\nobreakdash-$\Delta$ (low\nobreakdash-$\Delta$)
initial state. We then identify the region of non-zero $\Delta_{\textnormal{h}}-\Delta_{\textnormal{l}}$
as the regime where $\Delta$ (and hence $T_{e}$) is bistable. This
method is approximate mainly because the lifetimes of the metastable
states may be short compared to $5\textnormal{ ms}$. 

Figure~\ref{fig:phase-diag}(b) shows the theoretical prediction
for the bistable region in white. We generate it by numerically finding
the stationary solutions of Eq.~(\ref{eq:dDeltadt}), with $\Gamma\left(\Delta,\omega\right)$
determined from the fits shown in Fig.~\ref{fig:delta-calibration}(a)
\cite{Note1}. The qualitative features of the prediction agree well
with the experimental results. Quantitatively, the measured bistable
regime broadens in frequency faster than the predicted one. This discrepancy
is most likely due to  imperfect impedance matching of the probe
line and the imperfect correspondence between bistability and $\Delta_{\textnormal{h}}-\Delta_{\textnormal{l}}\ne0$.

The non-white areas in Fig.~\ref{fig:phase-diag}(b) show the prediction
for the susceptibility of the stationary-state $\Delta$ to external
heating, i.e., $\chi=\left.\partial\Delta/\partial P_{\textnormal{h}}\right|_{\dot{\Delta}=0}.$
It is a convenient dimensionless way to quantify the importance of
the electrothermal nonlinearity. Besides characterizing susceptibility
to heating, $\chi$ also gives the ratio of the effective thermal
time constant to its linear response value. Figure~\ref{fig:phase-diag}(b)
shows that both positive ($\chi\gg1$) and negative ($\chi\ll1$)
feedback regimes are accessible by simply choosing different values
of $f_{\textnormal{p}}$ and $P_{\textnormal{p}}$.

There are two distinct ways to operate the device as a detector in
the nonlinear regime. Approaching the bistable regime along line A
in Fig.~\ref{fig:phase-diag}, the system undergoes a pitchfork bifurcation
preceded by a diverging $\chi$. Analogously to the linear amplification
of coherent pulses by a Josephson parametric amplifier \cite{Zimmer1967PARAMETRIC,CastellanosBeltran2007Widely},
our device could in principle detect heat pulses in a continuous and
energy-resolving manner in this regime preceding the bifurcation.
However, the focus of this paper is threshold detection, which uses
the imperfect pitchfork bifurcation encountered along line B in Fig.~\ref{fig:phase-diag}
and bears a closer resemblance to the Josephson bifurcation amplifier
(JBA) \cite{Siddiqi2004RFDriven,Vijay2009Invited}.

\begin{figure}[!t]
\begin{centering}
\includegraphics{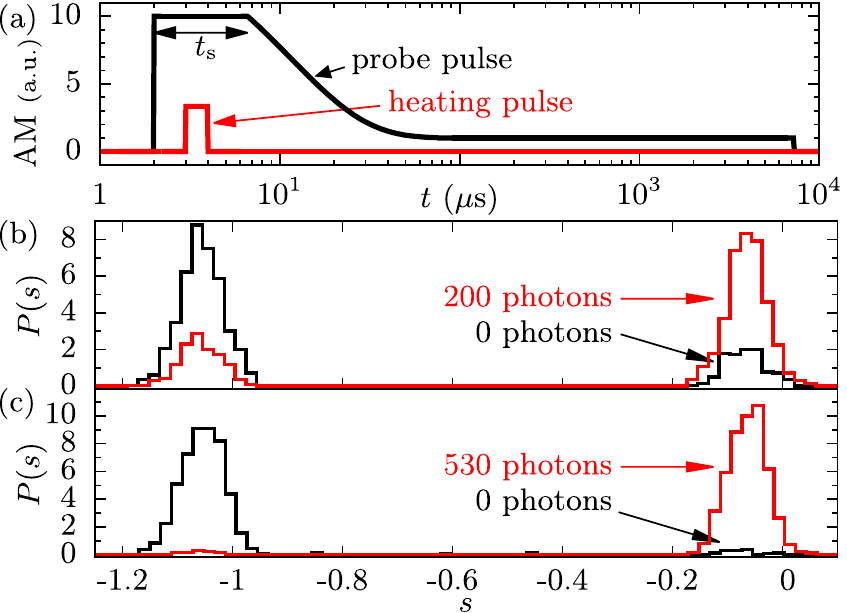}
\par\end{centering}

\caption{(Color online) (a) Amplitude modulation (AM) of the probe pulse used
for detecting weak $1\,\mu\textnormal{s}$ heating pulses (also shown).
The carrier frequencies are $757\textnormal{ MHz}$ and $8.4\textnormal{ GHz}$
for the probe and heating pulses, respectively.   (b) Normalized
histograms of the single-shot measurement outcome $s$ with a heating
pulse energy of zero or $200\times h\times8.4\,\textnormal{GHz}\approx1.1\textnormal{ zJ}$.
 The pulses for the two histograms were interleaved in time. (c)
Same as (b) but for $3.0\textnormal{ zJ}$ and $t_{\textnormal{s}}=2.5\textnormal{ \ensuremath{\mu}s}$.
\label{fig:hist} }
\end{figure}

In the threshold detection mode, we modulate the probe signal amplitude
as shown in Fig.~\ref{fig:hist}(a) while keeping the probe frequency
fixed at $f_{\textnormal{p}}=757\textnormal{ MHz}$. The amplitude
modulation pattern first initializes the system to a low\nobreakdash-$\Delta$
state, then makes it sensitive to a transition to the high\nobreakdash-$\Delta$
state for roughly $t_{\textnormal{s}}\approx4.5\textnormal{ \ensuremath{\mu}s}$
during which the heating pulse is sent, and finally keeps the system
in a long-lifetime part of the bistable regime for another $7\textnormal{ ms}$
in order to time-average the output signal. During the last stage
$P_{\textnormal{p}}\approx-131\textnormal{ dBm}$.  This is similar
to how JBAs operate \cite{Vijay2009Invited}. Note, however, that
the probe and heater signals do not interfere coherently due to the
transduction through electron temperature. That is, at heater frequencies
well above $\tau^{-1}$, the output signal is independent of the phase
of the heater signal. 

The histograms in Fig.~\ref{fig:hist}(b,c) show that the detector
switches reliably to the high-temperature state with a heating pulse
energy $E_{\textnormal{pulse}}\apprge1\textnormal{ zJ}$, while it
typically remains in the low-temperature state if no heating is applied.
The histograms are plotted against $s=\textnormal{Re}\left[e^{1.482\pi i}\int_{0.8\textnormal{ ms}}^{6.4\textnormal{ ms}}dt\Gamma\left(t\right)/\left(5.6\textnormal{ ms}\right)\right]$,
i.e., a projection of the time-averaged reflection coefficient. 
Few switching events occur during the averaging time, as indicated
by the scarcity of points between the two main peaks in the probability
density $P\left(s\right)$.  Instead, the errors arise from spurious
early switching events and events where the detector does not switch
despite a heating pulse. In particular, for a heat pulse of approximately
200 photons,  the readout fidelity is $F=0.56$ {[}Fig.~\ref{fig:hist}(b){]}.
Here, $F=1-P\left(s>-0.25|\textnormal{no heat pulse}\right)-P\left(s\leq-0.25|\textnormal{heat pulse}\right)$.
For a heat pulse of 530 photons, $F=0.94$ {[}Fig.~\ref{fig:hist}(c){]}.
For 330 photons, $F=0.75$~\cite{Note1}.

The observed pulse energy dependence of $F$ is in agreement with
the errors arising mainly from Gaussian fluctuations in the energy
of the nanowire electrons. Such fluctuations limit $F$ to $\bar{F}=\textnormal{erf}(2^{-3/2}E_{\textnormal{pulse}}/\Delta E_{\textnormal{RMS}})$,
even for ideal instantaneous threshold detection. For RMS fluctuations
$\Delta E_{\textnormal{RMS}}=0.7\textnormal{ zJ}$, $\bar{F}$ agrees
well with the above mentioned values of $F$. This phenomenological
$\Delta E_{\textnormal{RMS}}$ should be compared to the thermodynamic
fluctuations $\Delta E_{\textnormal{RMS}}^{\prime}=\sqrt{k_{\textnormal{B}}T_{e}^{2}C_{e}}$
in the absence of electrothermal feedback \cite{Moseley1984Thermal}.
For $T_{e}=50\textnormal{ mK}$, we estimate $C_{e}\approx400\textnormal{ zJ/K}$,
leading to $\Delta E_{\textnormal{RMS}}^{\prime}\approx0.12\textnormal{ zJ}\approx C_{e}\times0.29\textnormal{ mK}$.
This suggests that the thermodynamic fluctuations are a significant,
even if not the dominant, fidelity-limiting factor. Note that, although
the feedback during the pulse sequence in Fig.~\ref{fig:hist}(a)
is strong and positive, all signals are kept off for at least $400\textnormal{ ms}$
before each probe pulse. Therefore, the fluctuations just before the
brief period relevant for switching ($t_{\textnormal{s}}\ll\tau$)
are not affected by the electrothermal feedback.

In conclusion, we have experimentally investigated the electrothermal
feedback effect in a microwave photodetector. The results are in agreement
with a simple model which we used to highlight that both strong positive
and strong negative feedback is available by adjusting the probe power
and frequency. We demonstrated that bistability emerges in the limit
of extreme positive feedback and that it can be used for efficient
threshold detection of weak microwave pulses at the zeptojoule level.
This is more than an order of magnitude improvement over previous
thermal detector results, and therefore an important step toward thermal
detection of individual itinerant microwave photons.  To reach the
single-photon level, we should further reduce the nanowire volume
and possibly replace $\textnormal{Au}_{x}\textnormal{Pd}_{1-x}$ by
a material with lower specific heat. This would reduce the time constant
as well as the thermodynamic energy fluctuations, which contribute
significantly to the achieved fidelities according to our estimate.
Furthermore, there seems to be room for technical improvement in shielding
and filtering, which would bring the observed $\Delta E_{\textnormal{RMS}}$
closer to the thermodynamic fluctuations and would, most likely, lead
to a lower electron temperature.  Finally, a state-of-the-art amplifier
\cite{Vijay2011Observation,Andre1999Radiofrequency,Macklin2015Nearquantumlimited}
on the probe output should reduce the required averaging time by at
least two orders of magnitude \cite{Note1}.

\begin{acknowledgments}
We thank Leif Grönberg for depositing the Nb used in this work, Ari-Pekka
Soikkeli for discussion on modeling the bistability, and Matti Partanen
for technical assistance. We also acknowledge the financial support
from the Emil Aaltonen Foundation, the European Research Council under
Grant 278117 (``SINGLEOUT''), the Academy of Finland under Grants
265675, 251748, 284621, 135794, 272806, 286215, and 276528, and the
European Metrology Research Programme (``EXL03 MICROPHOTON''). The
EMRP is jointly funded by the EMRP participating countries within
EURAMET and the European Union. In addition, we acknowledge the provision
of facilities by Aalto University at OtaNano - Micronova Nanofabrication
Centre.
\end{acknowledgments}
\bibliography{govenius_tweaked}

\begin{thebibliography}{66}%
\makeatletter
\providecommand \@ifxundefined [1]{%
 \@ifx{#1\undefined}
}%
\providecommand \@ifnum [1]{%
 \ifnum #1\expandafter \@firstoftwo
 \else \expandafter \@secondoftwo
 \fi
}%
\providecommand \@ifx [1]{%
 \ifx #1\expandafter \@firstoftwo
 \else \expandafter \@secondoftwo
 \fi
}%
\providecommand \natexlab [1]{#1}%
\providecommand \enquote  [1]{``#1''}%
\providecommand \bibnamefont  [1]{#1}%
\providecommand \bibfnamefont [1]{#1}%
\providecommand \citenamefont [1]{#1}%
\providecommand \href@noop [0]{\@secondoftwo}%
\providecommand \href [0]{\begingroup \@sanitize@url \@href}%
\providecommand \@href[1]{\@@startlink{#1}\@@href}%
\providecommand \@@href[1]{\endgroup#1\@@endlink}%
\providecommand \@sanitize@url [0]{\catcode `\\12\catcode `\$12\catcode
  `\&12\catcode `\#12\catcode `\^12\catcode `\_12\catcode `\%12\relax}%
\providecommand \@@startlink[1]{}%
\providecommand \@@endlink[0]{}%
\providecommand \url  [0]{\begingroup\@sanitize@url \@url }%
\providecommand \@url [1]{\endgroup\@href {#1}{\urlprefix }}%
\providecommand \urlprefix  [0]{URL }%
\providecommand \Eprint [0]{\href }%
\providecommand \doibase [0]{http://dx.doi.org/}%
\providecommand \selectlanguage [0]{\@gobble}%
\providecommand \bibinfo  [0]{\@secondoftwo}%
\providecommand \bibfield  [0]{\@secondoftwo}%
\providecommand \translation [1]{[#1]}%
\providecommand \BibitemOpen [0]{}%
\providecommand \bibitemStop [0]{}%
\providecommand \bibitemNoStop [0]{.\EOS\space}%
\providecommand \EOS [0]{\spacefactor3000\relax}%
\providecommand \BibitemShut  [1]{\csname bibitem#1\endcsname}%
\let\auto@bib@innerbib\@empty
\bibitem [{\citenamefont {Blais}\ \emph {et~al.}(2004)\citenamefont {Blais},
  \citenamefont {Huang}, \citenamefont {Wallraff}, \citenamefont {Girvin},\
  and\ \citenamefont {Schoelkopf}}]{Blais04}%
  \BibitemOpen
  \bibfield  {author} {\bibinfo {author} {\bibfnamefont {A.}~\bibnamefont
  {Blais}}, \bibinfo {author} {\bibfnamefont {R.-S.}\ \bibnamefont {Huang}},
  \bibinfo {author} {\bibfnamefont {A.}~\bibnamefont {Wallraff}}, \bibinfo
  {author} {\bibfnamefont {S.~M.}\ \bibnamefont {Girvin}}, \ and\ \bibinfo
  {author} {\bibfnamefont {R.~J.}\ \bibnamefont {Schoelkopf}},\ }\href
  {\doibase 10.1103/physreva.69.062320} {\bibfield  {journal} {\bibinfo
  {journal} {Phys. Rev. A}\ }\textbf {\bibinfo {volume} {69}},\ \bibinfo
  {pages} {062320} (\bibinfo {year} {2004})}\BibitemShut {NoStop}%
\bibitem [{\citenamefont {Wallraff}\ \emph {et~al.}(2004)\citenamefont
  {Wallraff}, \citenamefont {Schuster}, \citenamefont {Blais}, \citenamefont
  {Frunzio}, \citenamefont {Huang}, \citenamefont {Majer}, \citenamefont
  {Kumar}, \citenamefont {Girvin},\ and\ \citenamefont
  {Schoelkopf}}]{Wallraff2004Strong}%
  \BibitemOpen
  \bibfield  {author} {\bibinfo {author} {\bibfnamefont {A.}~\bibnamefont
  {Wallraff}}, \bibinfo {author} {\bibfnamefont {D.~I.}\ \bibnamefont
  {Schuster}}, \bibinfo {author} {\bibfnamefont {A.}~\bibnamefont {Blais}},
  \bibinfo {author} {\bibfnamefont {L.}~\bibnamefont {Frunzio}}, \bibinfo
  {author} {\bibfnamefont {R.-S.}\ \bibnamefont {Huang}}, \bibinfo {author}
  {\bibfnamefont {J.}~\bibnamefont {Majer}}, \bibinfo {author} {\bibfnamefont
  {S.}~\bibnamefont {Kumar}}, \bibinfo {author} {\bibfnamefont {S.~M.}\
  \bibnamefont {Girvin}}, \ and\ \bibinfo {author} {\bibfnamefont {R.~J.}\
  \bibnamefont {Schoelkopf}},\ }\href {\doibase 10.1038/nature02851} {\bibfield
   {journal} {\bibinfo  {journal} {Nature (London)}\ }\textbf {\bibinfo
  {volume} {431}},\ \bibinfo {pages} {162} (\bibinfo {year}
  {2004})}\BibitemShut {NoStop}%
\bibitem [{\citenamefont {Devoret}\ and\ \citenamefont
  {Schoelkopf}(2013)}]{Devoret2013Superconducting}%
  \BibitemOpen
  \bibfield  {author} {\bibinfo {author} {\bibfnamefont {M.~H.}\ \bibnamefont
  {Devoret}}\ and\ \bibinfo {author} {\bibfnamefont {R.~J.}\ \bibnamefont
  {Schoelkopf}},\ }\href {\doibase 10.1126/science.1231930} {\bibfield
  {journal} {\bibinfo  {journal} {Science}\ }\textbf {\bibinfo {volume}
  {339}},\ \bibinfo {pages} {1169} (\bibinfo {year} {2013})}\BibitemShut
  {NoStop}%
\bibitem [{\citenamefont {Kelly}\ \emph {et~al.}(2015)\citenamefont {Kelly},
  \citenamefont {Barends}, \citenamefont {Fowler}, \citenamefont {Megrant},
  \citenamefont {Jeffrey}, \citenamefont {White}, \citenamefont {Sank},
  \citenamefont {Mutus}, \citenamefont {Campbell}, \citenamefont {Chen},
  \citenamefont {Chen}, \citenamefont {Chiaro}, \citenamefont {Dunsworth},
  \citenamefont {Hoi}, \citenamefont {Neill}, \citenamefont {O'Malley},
  \citenamefont {Quintana}, \citenamefont {Roushan}, \citenamefont
  {Vainsencher}, \citenamefont {Wenner}, \citenamefont {Cleland},\ and\
  \citenamefont {Martinis}}]{Kelly2015State}%
  \BibitemOpen
  \bibfield  {author} {\bibinfo {author} {\bibfnamefont {J.}~\bibnamefont
  {Kelly}}, \bibinfo {author} {\bibfnamefont {R.}~\bibnamefont {Barends}},
  \bibinfo {author} {\bibfnamefont {A.~G.}\ \bibnamefont {Fowler}}, \bibinfo
  {author} {\bibfnamefont {A.}~\bibnamefont {Megrant}}, \bibinfo {author}
  {\bibfnamefont {E.}~\bibnamefont {Jeffrey}}, \bibinfo {author} {\bibfnamefont
  {T.~C.}\ \bibnamefont {White}}, \bibinfo {author} {\bibfnamefont
  {D.}~\bibnamefont {Sank}}, \bibinfo {author} {\bibfnamefont {J.~Y.}\
  \bibnamefont {Mutus}}, \bibinfo {author} {\bibfnamefont {B.}~\bibnamefont
  {Campbell}}, \bibinfo {author} {\bibfnamefont {Y.}~\bibnamefont {Chen}},
  \bibinfo {author} {\bibfnamefont {Z.}~\bibnamefont {Chen}}, \bibinfo {author}
  {\bibfnamefont {B.}~\bibnamefont {Chiaro}}, \bibinfo {author} {\bibfnamefont
  {A.}~\bibnamefont {Dunsworth}}, \bibinfo {author} {\bibfnamefont {I.~C.}\
  \bibnamefont {Hoi}}, \bibinfo {author} {\bibfnamefont {C.}~\bibnamefont
  {Neill}}, \bibinfo {author} {\bibfnamefont {P.~J.~J.}\ \bibnamefont
  {O'Malley}}, \bibinfo {author} {\bibfnamefont {C.}~\bibnamefont {Quintana}},
  \bibinfo {author} {\bibfnamefont {P.}~\bibnamefont {Roushan}}, \bibinfo
  {author} {\bibfnamefont {A.}~\bibnamefont {Vainsencher}}, \bibinfo {author}
  {\bibfnamefont {J.}~\bibnamefont {Wenner}}, \bibinfo {author} {\bibfnamefont
  {A.~N.}\ \bibnamefont {Cleland}}, \ and\ \bibinfo {author} {\bibfnamefont
  {J.~M.}\ \bibnamefont {Martinis}},\ }\href {\doibase 10.1038/nature14270}
  {\bibfield  {journal} {\bibinfo  {journal} {Nature (London)}\ }\textbf
  {\bibinfo {volume} {519}},\ \bibinfo {pages} {66} (\bibinfo {year}
  {2015})}\BibitemShut {NoStop}%
\bibitem [{\citenamefont {Lita}\ \emph {et~al.}(2008)\citenamefont {Lita},
  \citenamefont {Miller},\ and\ \citenamefont {Nam}}]{Lita2008Counting}%
  \BibitemOpen
  \bibfield  {author} {\bibinfo {author} {\bibfnamefont {A.~E.}\ \bibnamefont
  {Lita}}, \bibinfo {author} {\bibfnamefont {A.~J.}\ \bibnamefont {Miller}}, \
  and\ \bibinfo {author} {\bibfnamefont {S.~W.}\ \bibnamefont {Nam}},\ }\href
  {\doibase 10.1364/oe.16.003032} {\bibfield  {journal} {\bibinfo  {journal}
  {Opt. Express}\ }\textbf {\bibinfo {volume} {16}},\ \bibinfo {pages} {3032}
  (\bibinfo {year} {2008})}\BibitemShut {NoStop}%
\bibitem [{\citenamefont {Marsili}\ \emph {et~al.}(2013)\citenamefont
  {Marsili}, \citenamefont {Verma}, \citenamefont {Stern}, \citenamefont
  {Harrington}, \citenamefont {Lita}, \citenamefont {Gerrits}, \citenamefont
  {Vayshenker}, \citenamefont {Baek}, \citenamefont {Shaw}, \citenamefont
  {Mirin},\ and\ \citenamefont {Nam}}]{Marsili2013Detecting}%
  \BibitemOpen
  \bibfield  {author} {\bibinfo {author} {\bibfnamefont {F.}~\bibnamefont
  {Marsili}}, \bibinfo {author} {\bibfnamefont {V.~B.}\ \bibnamefont {Verma}},
  \bibinfo {author} {\bibfnamefont {J.~A.}\ \bibnamefont {Stern}}, \bibinfo
  {author} {\bibfnamefont {S.}~\bibnamefont {Harrington}}, \bibinfo {author}
  {\bibfnamefont {A.~E.}\ \bibnamefont {Lita}}, \bibinfo {author}
  {\bibfnamefont {T.}~\bibnamefont {Gerrits}}, \bibinfo {author} {\bibfnamefont
  {I.}~\bibnamefont {Vayshenker}}, \bibinfo {author} {\bibfnamefont
  {B.}~\bibnamefont {Baek}}, \bibinfo {author} {\bibfnamefont {M.~D.}\
  \bibnamefont {Shaw}}, \bibinfo {author} {\bibfnamefont {R.~P.}\ \bibnamefont
  {Mirin}}, \ and\ \bibinfo {author} {\bibfnamefont {S.~W.}\ \bibnamefont
  {Nam}},\ }\href {\doibase 10.1038/nphoton.2013.13} {\bibfield  {journal}
  {\bibinfo  {journal} {Nat. Photonics}\ }\textbf {\bibinfo {volume} {7}},\
  \bibinfo {pages} {210} (\bibinfo {year} {2013})}\BibitemShut {NoStop}%
\bibitem [{\citenamefont {Eisaman}\ \emph {et~al.}(2011)\citenamefont
  {Eisaman}, \citenamefont {Fan}, \citenamefont {Migdall},\ and\ \citenamefont
  {Polyakov}}]{Eisaman2011Invited}%
  \BibitemOpen
  \bibfield  {author} {\bibinfo {author} {\bibfnamefont {M.~D.}\ \bibnamefont
  {Eisaman}}, \bibinfo {author} {\bibfnamefont {J.}~\bibnamefont {Fan}},
  \bibinfo {author} {\bibfnamefont {A.}~\bibnamefont {Migdall}}, \ and\
  \bibinfo {author} {\bibfnamefont {S.~V.}\ \bibnamefont {Polyakov}},\ }\href
  {\doibase 10.1063/1.3610677} {\bibfield  {journal} {\bibinfo  {journal} {Rev.
  Sci. Instrum.}\ }\textbf {\bibinfo {volume} {82}},\ \bibinfo {pages} {071101}
  (\bibinfo {year} {2011})}\BibitemShut {NoStop}%
\bibitem [{\citenamefont {Kok}\ \emph {et~al.}(2007)\citenamefont {Kok},
  \citenamefont {Nemoto}, \citenamefont {Ralph}, \citenamefont {Dowling},\ and\
  \citenamefont {Milburn}}]{Kok2007Linear}%
  \BibitemOpen
  \bibfield  {author} {\bibinfo {author} {\bibfnamefont {P.}~\bibnamefont
  {Kok}}, \bibinfo {author} {\bibfnamefont {K.}~\bibnamefont {Nemoto}},
  \bibinfo {author} {\bibfnamefont {T.~C.}\ \bibnamefont {Ralph}}, \bibinfo
  {author} {\bibfnamefont {J.~P.}\ \bibnamefont {Dowling}}, \ and\ \bibinfo
  {author} {\bibfnamefont {G.~J.}\ \bibnamefont {Milburn}},\ }\href {\doibase
  10.1103/revmodphys.79.135} {\bibfield  {journal} {\bibinfo  {journal} {Rev.
  Mod. Phys.}\ }\textbf {\bibinfo {volume} {79}},\ \bibinfo {pages} {135}
  (\bibinfo {year} {2007})}\BibitemShut {NoStop}%
\bibitem [{\citenamefont {Hillery}\ and\ \citenamefont
  {Scully}(1983)}]{Hillery1983State}%
  \BibitemOpen
  \bibfield  {author} {\bibinfo {author} {\bibfnamefont {M.}~\bibnamefont
  {Hillery}}\ and\ \bibinfo {author} {\bibfnamefont {M.~O.}\ \bibnamefont
  {Scully}},\ }in\ \href {\doibase 10.1007/978-1-4613-3712-6\_4} {\emph
  {\bibinfo {booktitle} {Quantum Optics, Experimental Gravity, and Measurement
  Theory}}},\ \bibinfo {series} {NATO Advanced Science Institutes Series},
  Vol.~\bibinfo {volume} {94},\ \bibinfo {editor} {edited by\ \bibinfo {editor}
  {\bibfnamefont {P.}~\bibnamefont {Meystre}}\ and\ \bibinfo {editor}
  {\bibfnamefont {M.~O.}\ \bibnamefont {Scully}}}\ (\bibinfo  {publisher}
  {Plenum Press},\ \bibinfo {address} {New York},\ \bibinfo {year} {1983})\
  pp.\ \bibinfo {pages} {65--85}\BibitemShut {NoStop}%
\bibitem [{\citenamefont {Govenius}\ \emph {et~al.}(2015)\citenamefont
  {Govenius}, \citenamefont {Matsuzaki}, \citenamefont {Savenko},\ and\
  \citenamefont {M\"{o}tt\"{o}nen}}]{Govenius2015Parity}%
  \BibitemOpen
  \bibfield  {author} {\bibinfo {author} {\bibfnamefont {J.}~\bibnamefont
  {Govenius}}, \bibinfo {author} {\bibfnamefont {Y.}~\bibnamefont {Matsuzaki}},
  \bibinfo {author} {\bibfnamefont {I.~G.}\ \bibnamefont {Savenko}}, \ and\
  \bibinfo {author} {\bibfnamefont {M.}~\bibnamefont {M\"{o}tt\"{o}nen}},\
  }\href {\doibase 10.1103/physreva.92.042305} {\bibfield  {journal} {\bibinfo
  {journal} {Phys. Rev. A}\ }\textbf {\bibinfo {volume} {92}},\ \bibinfo
  {pages} {042305} (\bibinfo {year} {2015})}\BibitemShut {NoStop}%
\bibitem [{\citenamefont {da~Silva}\ \emph {et~al.}(2010)\citenamefont
  {da~Silva}, \citenamefont {Bozyigit}, \citenamefont {Wallraff},\ and\
  \citenamefont {Blais}}]{daSilva2010Schemes}%
  \BibitemOpen
  \bibfield  {author} {\bibinfo {author} {\bibfnamefont {M.~P.}\ \bibnamefont
  {da~Silva}}, \bibinfo {author} {\bibfnamefont {D.}~\bibnamefont {Bozyigit}},
  \bibinfo {author} {\bibfnamefont {A.}~\bibnamefont {Wallraff}}, \ and\
  \bibinfo {author} {\bibfnamefont {A.}~\bibnamefont {Blais}},\ }\href
  {\doibase 10.1103/physreva.82.043804} {\bibfield  {journal} {\bibinfo
  {journal} {Phys. Rev. A}\ }\textbf {\bibinfo {volume} {82}},\ \bibinfo
  {pages} {043804} (\bibinfo {year} {2010})}\BibitemShut {NoStop}%
\bibitem [{\citenamefont {Bozyigit}\ \emph {et~al.}(2010)\citenamefont
  {Bozyigit}, \citenamefont {Lang}, \citenamefont {Steffen}, \citenamefont
  {Fink}, \citenamefont {Eichler}, \citenamefont {Baur}, \citenamefont
  {Bianchetti}, \citenamefont {Leek}, \citenamefont {Filipp}, \citenamefont
  {da~Silva}, \citenamefont {Blais},\ and\ \citenamefont
  {Wallraff}}]{Bozyigit2010Antibunching}%
  \BibitemOpen
  \bibfield  {author} {\bibinfo {author} {\bibfnamefont {D.}~\bibnamefont
  {Bozyigit}}, \bibinfo {author} {\bibfnamefont {C.}~\bibnamefont {Lang}},
  \bibinfo {author} {\bibfnamefont {L.}~\bibnamefont {Steffen}}, \bibinfo
  {author} {\bibfnamefont {J.~M.}\ \bibnamefont {Fink}}, \bibinfo {author}
  {\bibfnamefont {C.}~\bibnamefont {Eichler}}, \bibinfo {author} {\bibfnamefont
  {M.}~\bibnamefont {Baur}}, \bibinfo {author} {\bibfnamefont {R.}~\bibnamefont
  {Bianchetti}}, \bibinfo {author} {\bibfnamefont {P.~J.}\ \bibnamefont
  {Leek}}, \bibinfo {author} {\bibfnamefont {S.}~\bibnamefont {Filipp}},
  \bibinfo {author} {\bibfnamefont {M.~P.}\ \bibnamefont {da~Silva}}, \bibinfo
  {author} {\bibfnamefont {A.}~\bibnamefont {Blais}}, \ and\ \bibinfo {author}
  {\bibfnamefont {A.}~\bibnamefont {Wallraff}},\ }\href {\doibase
  10.1038/nphys1845} {\bibfield  {journal} {\bibinfo  {journal} {Nature Phys.}\
  }\textbf {\bibinfo {volume} {7}},\ \bibinfo {pages} {154} (\bibinfo {year}
  {2010})}\BibitemShut {NoStop}%
\bibitem [{\citenamefont {Karasik}\ \emph {et~al.}(2011)\citenamefont
  {Karasik}, \citenamefont {Sergeev},\ and\ \citenamefont
  {Prober}}]{Karasik2011Nanobolometers}%
  \BibitemOpen
  \bibfield  {author} {\bibinfo {author} {\bibfnamefont {B.~S.}\ \bibnamefont
  {Karasik}}, \bibinfo {author} {\bibfnamefont {A.~V.}\ \bibnamefont
  {Sergeev}}, \ and\ \bibinfo {author} {\bibfnamefont {D.~E.}\ \bibnamefont
  {Prober}},\ }\href {\doibase 10.1109/tthz.2011.2159560} {\bibfield  {journal}
  {\bibinfo  {journal} {IEEE Trans. Terahertz Sci.}\ }\textbf {\bibinfo
  {volume} {1}},\ \bibinfo {pages} {97} (\bibinfo {year} {2011})}\BibitemShut
  {NoStop}%
\bibitem [{\citenamefont {Karasik}\ \emph {et~al.}(2012)\citenamefont
  {Karasik}, \citenamefont {Pereverzev}, \citenamefont {Soibel}, \citenamefont
  {Santavicca}, \citenamefont {Prober}, \citenamefont {Olaya},\ and\
  \citenamefont {Gershenson}}]{Karasik2012Energyresolved}%
  \BibitemOpen
  \bibfield  {author} {\bibinfo {author} {\bibfnamefont {B.~S.}\ \bibnamefont
  {Karasik}}, \bibinfo {author} {\bibfnamefont {S.~V.}\ \bibnamefont
  {Pereverzev}}, \bibinfo {author} {\bibfnamefont {A.}~\bibnamefont {Soibel}},
  \bibinfo {author} {\bibfnamefont {D.~F.}\ \bibnamefont {Santavicca}},
  \bibinfo {author} {\bibfnamefont {D.~E.}\ \bibnamefont {Prober}}, \bibinfo
  {author} {\bibfnamefont {D.}~\bibnamefont {Olaya}}, \ and\ \bibinfo {author}
  {\bibfnamefont {M.~E.}\ \bibnamefont {Gershenson}},\ }\href {\doibase
  10.1063/1.4739839} {\bibfield  {journal} {\bibinfo  {journal} {Appl. Phys.
  Lett.}\ }\textbf {\bibinfo {volume} {101}},\ \bibinfo {pages} {052601}
  (\bibinfo {year} {2012})}\BibitemShut {NoStop}%
\bibitem [{\citenamefont {Pekola}(2015)}]{Pekola2015Towards}%
  \BibitemOpen
  \bibfield  {author} {\bibinfo {author} {\bibfnamefont {J.~P.}\ \bibnamefont
  {Pekola}},\ }\href {\doibase 10.1038/nphys3169} {\bibfield  {journal}
  {\bibinfo  {journal} {Nature Phys.}\ }\textbf {\bibinfo {volume} {11}},\
  \bibinfo {pages} {118} (\bibinfo {year} {2015})}\BibitemShut {NoStop}%
\bibitem [{\citenamefont {Pekola}\ \emph {et~al.}(2013)\citenamefont {Pekola},
  \citenamefont {Solinas}, \citenamefont {Shnirman},\ and\ \citenamefont
  {Averin}}]{Pekola2013Calorimetric}%
  \BibitemOpen
  \bibfield  {author} {\bibinfo {author} {\bibfnamefont {J.~P.}\ \bibnamefont
  {Pekola}}, \bibinfo {author} {\bibfnamefont {P.}~\bibnamefont {Solinas}},
  \bibinfo {author} {\bibfnamefont {A.}~\bibnamefont {Shnirman}}, \ and\
  \bibinfo {author} {\bibfnamefont {D.~V.}\ \bibnamefont {Averin}},\ }\href
  {\doibase 10.1088/1367-2630/15/11/115006} {\bibfield  {journal} {\bibinfo
  {journal} {New J. Phys.}\ }\textbf {\bibinfo {volume} {15}},\ \bibinfo
  {pages} {115006} (\bibinfo {year} {2013})}\BibitemShut {NoStop}%
\bibitem [{\citenamefont {Schmidt}\ \emph {et~al.}(2003)\citenamefont
  {Schmidt}, \citenamefont {Yung},\ and\ \citenamefont
  {Cleland}}]{Schmidt2003Nanoscale}%
  \BibitemOpen
  \bibfield  {author} {\bibinfo {author} {\bibfnamefont {D.~R.}\ \bibnamefont
  {Schmidt}}, \bibinfo {author} {\bibfnamefont {C.~S.}\ \bibnamefont {Yung}}, \
  and\ \bibinfo {author} {\bibfnamefont {A.~N.}\ \bibnamefont {Cleland}},\
  }\href {\doibase 10.1063/1.1597983} {\bibfield  {journal} {\bibinfo
  {journal} {Appl. Phys. Lett.}\ }\textbf {\bibinfo {volume} {83}},\ \bibinfo
  {pages} {1002} (\bibinfo {year} {2003})}\BibitemShut {NoStop}%
\bibitem [{\citenamefont {Schmidt}\ \emph {et~al.}(2005)\citenamefont
  {Schmidt}, \citenamefont {Lehnert}, \citenamefont {Clark}, \citenamefont
  {Duncan}, \citenamefont {Irwin}, \citenamefont {Miller},\ and\ \citenamefont
  {Ullom}}]{Schmidt2005Superconductorinsulatornormal}%
  \BibitemOpen
  \bibfield  {author} {\bibinfo {author} {\bibfnamefont {D.~R.}\ \bibnamefont
  {Schmidt}}, \bibinfo {author} {\bibfnamefont {K.~W.}\ \bibnamefont
  {Lehnert}}, \bibinfo {author} {\bibfnamefont {A.~M.}\ \bibnamefont {Clark}},
  \bibinfo {author} {\bibfnamefont {W.~D.}\ \bibnamefont {Duncan}}, \bibinfo
  {author} {\bibfnamefont {K.~D.}\ \bibnamefont {Irwin}}, \bibinfo {author}
  {\bibfnamefont {N.}~\bibnamefont {Miller}}, \ and\ \bibinfo {author}
  {\bibfnamefont {J.~N.}\ \bibnamefont {Ullom}},\ }\href {\doibase
  10.1063/1.1855411} {\bibfield  {journal} {\bibinfo  {journal} {Appl. Phys.
  Lett.}\ }\textbf {\bibinfo {volume} {86}},\ \bibinfo {pages} {053505}
  (\bibinfo {year} {2005})}\BibitemShut {NoStop}%
\bibitem [{\citenamefont {Gasparinetti}\ \emph {et~al.}(2015)\citenamefont
  {Gasparinetti}, \citenamefont {Viisanen}, \citenamefont {Saira},
  \citenamefont {Faivre}, \citenamefont {Arzeo}, \citenamefont {Meschke},\ and\
  \citenamefont {Pekola}}]{Gasparinetti2015Fast}%
  \BibitemOpen
  \bibfield  {author} {\bibinfo {author} {\bibfnamefont {S.}~\bibnamefont
  {Gasparinetti}}, \bibinfo {author} {\bibfnamefont {K.~L.}\ \bibnamefont
  {Viisanen}}, \bibinfo {author} {\bibfnamefont {O.-P.}\ \bibnamefont {Saira}},
  \bibinfo {author} {\bibfnamefont {T.}~\bibnamefont {Faivre}}, \bibinfo
  {author} {\bibfnamefont {M.}~\bibnamefont {Arzeo}}, \bibinfo {author}
  {\bibfnamefont {M.}~\bibnamefont {Meschke}}, \ and\ \bibinfo {author}
  {\bibfnamefont {J.~P.}\ \bibnamefont {Pekola}},\ }\href {\doibase
  10.1103/physrevapplied.3.014007} {\bibfield  {journal} {\bibinfo  {journal}
  {Phys. Rev. Appl.}\ }\textbf {\bibinfo {volume} {3}},\ \bibinfo {pages}
  {014007} (\bibinfo {year} {2015})}\BibitemShut {NoStop}%
\bibitem [{\citenamefont {Saira}\ \emph {et~al.}()\citenamefont {Saira},
  \citenamefont {Zgirski}, \citenamefont {Viisanen}, \citenamefont {Golubev},\
  and\ \citenamefont {Pekola}}]{Saira2016Dispersive}%
  \BibitemOpen
  \bibfield  {author} {\bibinfo {author} {\bibfnamefont {O.~P.}\ \bibnamefont
  {Saira}}, \bibinfo {author} {\bibfnamefont {M.}~\bibnamefont {Zgirski}},
  \bibinfo {author} {\bibfnamefont {K.~L.}\ \bibnamefont {Viisanen}}, \bibinfo
  {author} {\bibfnamefont {D.~S.}\ \bibnamefont {Golubev}}, \ and\ \bibinfo
  {author} {\bibfnamefont {J.~P.}\ \bibnamefont {Pekola}},\ }\href@noop {}
  {}\Eprint {http://arxiv.org/abs/1604.05089} {arXiv:1604.05089} \BibitemShut
  {NoStop}%
\bibitem [{\citenamefont {Romero}\ \emph {et~al.}(2009)\citenamefont {Romero},
  \citenamefont {Garc\'{\i}a-Ripoll},\ and\ \citenamefont
  {Solano}}]{Romero2009Microwave}%
  \BibitemOpen
  \bibfield  {author} {\bibinfo {author} {\bibfnamefont {G.}~\bibnamefont
  {Romero}}, \bibinfo {author} {\bibfnamefont {J.~J.}\ \bibnamefont
  {Garc\'{\i}a-Ripoll}}, \ and\ \bibinfo {author} {\bibfnamefont
  {E.}~\bibnamefont {Solano}},\ }\href {\doibase
  10.1103/physrevlett.102.173602} {\bibfield  {journal} {\bibinfo  {journal}
  {Phys. Rev. Lett.}\ }\textbf {\bibinfo {volume} {102}},\ \bibinfo {pages}
  {173602} (\bibinfo {year} {2009})}\BibitemShut {NoStop}%
\bibitem [{\citenamefont {Chen}\ \emph {et~al.}(2011)\citenamefont {Chen},
  \citenamefont {Hover}, \citenamefont {Sendelbach}, \citenamefont {Maurer},
  \citenamefont {Merkel}, \citenamefont {Pritchett}, \citenamefont {Wilhelm},\
  and\ \citenamefont {McDermott}}]{Chen2011Microwave}%
  \BibitemOpen
  \bibfield  {author} {\bibinfo {author} {\bibfnamefont {Y.~F.}\ \bibnamefont
  {Chen}}, \bibinfo {author} {\bibfnamefont {D.}~\bibnamefont {Hover}},
  \bibinfo {author} {\bibfnamefont {S.}~\bibnamefont {Sendelbach}}, \bibinfo
  {author} {\bibfnamefont {L.}~\bibnamefont {Maurer}}, \bibinfo {author}
  {\bibfnamefont {S.~T.}\ \bibnamefont {Merkel}}, \bibinfo {author}
  {\bibfnamefont {E.~J.}\ \bibnamefont {Pritchett}}, \bibinfo {author}
  {\bibfnamefont {F.~K.}\ \bibnamefont {Wilhelm}}, \ and\ \bibinfo {author}
  {\bibfnamefont {R.}~\bibnamefont {McDermott}},\ }\href {\doibase
  10.1103/physrevlett.107.217401} {\bibfield  {journal} {\bibinfo  {journal}
  {Phys. Rev. Lett.}\ }\textbf {\bibinfo {volume} {107}},\ \bibinfo {pages}
  {217401} (\bibinfo {year} {2011})}\BibitemShut {NoStop}%
\bibitem [{\citenamefont {Peropadre}\ \emph {et~al.}(2011)\citenamefont
  {Peropadre}, \citenamefont {Romero}, \citenamefont {Johansson}, \citenamefont
  {Wilson}, \citenamefont {Solano},\ and\ \citenamefont
  {Garc\'{\i}a-Ripoll}}]{Peropadre2011Approaching}%
  \BibitemOpen
  \bibfield  {author} {\bibinfo {author} {\bibfnamefont {B.}~\bibnamefont
  {Peropadre}}, \bibinfo {author} {\bibfnamefont {G.}~\bibnamefont {Romero}},
  \bibinfo {author} {\bibfnamefont {G.}~\bibnamefont {Johansson}}, \bibinfo
  {author} {\bibfnamefont {C.~M.}\ \bibnamefont {Wilson}}, \bibinfo {author}
  {\bibfnamefont {E.}~\bibnamefont {Solano}}, \ and\ \bibinfo {author}
  {\bibfnamefont {J.~J.}\ \bibnamefont {Garc\'{\i}a-Ripoll}},\ }\href {\doibase
  10.1103/physreva.84.063834} {\bibfield  {journal} {\bibinfo  {journal} {Phys.
  Rev. A}\ }\textbf {\bibinfo {volume} {84}},\ \bibinfo {pages} {063834}
  (\bibinfo {year} {2011})}\BibitemShut {NoStop}%
\bibitem [{\citenamefont {Fan}\ \emph {et~al.}(2013)\citenamefont {Fan},
  \citenamefont {Kockum}, \citenamefont {Combes}, \citenamefont {Johansson},
  \citenamefont {Hoi}, \citenamefont {Wilson}, \citenamefont {Delsing},
  \citenamefont {Milburn},\ and\ \citenamefont {Stace}}]{Fan2013Breakdown}%
  \BibitemOpen
  \bibfield  {author} {\bibinfo {author} {\bibfnamefont {B.}~\bibnamefont
  {Fan}}, \bibinfo {author} {\bibfnamefont {A.~F.}\ \bibnamefont {Kockum}},
  \bibinfo {author} {\bibfnamefont {J.}~\bibnamefont {Combes}}, \bibinfo
  {author} {\bibfnamefont {G.}~\bibnamefont {Johansson}}, \bibinfo {author}
  {\bibfnamefont {I.-C.}\ \bibnamefont {Hoi}}, \bibinfo {author} {\bibfnamefont
  {C.~M.}\ \bibnamefont {Wilson}}, \bibinfo {author} {\bibfnamefont
  {P.}~\bibnamefont {Delsing}}, \bibinfo {author} {\bibfnamefont {G.~J.}\
  \bibnamefont {Milburn}}, \ and\ \bibinfo {author} {\bibfnamefont {T.~M.}\
  \bibnamefont {Stace}},\ }\href {\doibase 10.1103/physrevlett.110.053601}
  {\bibfield  {journal} {\bibinfo  {journal} {Phys. Rev. Lett.}\ }\textbf
  {\bibinfo {volume} {110}},\ \bibinfo {pages} {053601} (\bibinfo {year}
  {2013})}\BibitemShut {NoStop}%
\bibitem [{\citenamefont {Hoi}\ \emph {et~al.}(2013)\citenamefont {Hoi},
  \citenamefont {Kockum}, \citenamefont {Palomaki}, \citenamefont {Stace},
  \citenamefont {Fan}, \citenamefont {Tornberg}, \citenamefont {Sathyamoorthy},
  \citenamefont {Johansson}, \citenamefont {Delsing},\ and\ \citenamefont
  {Wilson}}]{Hoi2013Giant}%
  \BibitemOpen
  \bibfield  {author} {\bibinfo {author} {\bibfnamefont {I.-C.}\ \bibnamefont
  {Hoi}}, \bibinfo {author} {\bibfnamefont {A.~F.}\ \bibnamefont {Kockum}},
  \bibinfo {author} {\bibfnamefont {T.}~\bibnamefont {Palomaki}}, \bibinfo
  {author} {\bibfnamefont {T.~M.}\ \bibnamefont {Stace}}, \bibinfo {author}
  {\bibfnamefont {B.}~\bibnamefont {Fan}}, \bibinfo {author} {\bibfnamefont
  {L.}~\bibnamefont {Tornberg}}, \bibinfo {author} {\bibfnamefont {S.~R.}\
  \bibnamefont {Sathyamoorthy}}, \bibinfo {author} {\bibfnamefont
  {G.}~\bibnamefont {Johansson}}, \bibinfo {author} {\bibfnamefont
  {P.}~\bibnamefont {Delsing}}, \ and\ \bibinfo {author} {\bibfnamefont
  {C.~M.}\ \bibnamefont {Wilson}},\ }\href {\doibase
  10.1103/physrevlett.111.053601} {\bibfield  {journal} {\bibinfo  {journal}
  {Phys. Rev. Lett.}\ }\textbf {\bibinfo {volume} {111}},\ \bibinfo {pages}
  {053601} (\bibinfo {year} {2013})}\BibitemShut {NoStop}%
\bibitem [{\citenamefont {Sathyamoorthy}\ \emph {et~al.}(2014)\citenamefont
  {Sathyamoorthy}, \citenamefont {Tornberg}, \citenamefont {Kockum},
  \citenamefont {Baragiola}, \citenamefont {Combes}, \citenamefont {Wilson},
  \citenamefont {Stace},\ and\ \citenamefont
  {Johansson}}]{Sathyamoorthy2014Quantum}%
  \BibitemOpen
  \bibfield  {author} {\bibinfo {author} {\bibfnamefont {S.~R.}\ \bibnamefont
  {Sathyamoorthy}}, \bibinfo {author} {\bibfnamefont {L.}~\bibnamefont
  {Tornberg}}, \bibinfo {author} {\bibfnamefont {A.~F.}\ \bibnamefont
  {Kockum}}, \bibinfo {author} {\bibfnamefont {B.~Q.}\ \bibnamefont
  {Baragiola}}, \bibinfo {author} {\bibfnamefont {J.}~\bibnamefont {Combes}},
  \bibinfo {author} {\bibfnamefont {C.~M.}\ \bibnamefont {Wilson}}, \bibinfo
  {author} {\bibfnamefont {T.~M.}\ \bibnamefont {Stace}}, \ and\ \bibinfo
  {author} {\bibfnamefont {G.}~\bibnamefont {Johansson}},\ }\href {\doibase
  10.1103/physrevlett.112.093601} {\bibfield  {journal} {\bibinfo  {journal}
  {Phys. Rev. Lett.}\ }\textbf {\bibinfo {volume} {112}},\ \bibinfo {pages}
  {093601} (\bibinfo {year} {2014})}\BibitemShut {NoStop}%
\bibitem [{\citenamefont {Fan}\ \emph {et~al.}(2014)\citenamefont {Fan},
  \citenamefont {Johansson}, \citenamefont {Combes}, \citenamefont {Milburn},\
  and\ \citenamefont {Stace}}]{Fan2014Nonabsorbing}%
  \BibitemOpen
  \bibfield  {author} {\bibinfo {author} {\bibfnamefont {B.}~\bibnamefont
  {Fan}}, \bibinfo {author} {\bibfnamefont {G.}~\bibnamefont {Johansson}},
  \bibinfo {author} {\bibfnamefont {J.}~\bibnamefont {Combes}}, \bibinfo
  {author} {\bibfnamefont {G.~J.}\ \bibnamefont {Milburn}}, \ and\ \bibinfo
  {author} {\bibfnamefont {T.~M.}\ \bibnamefont {Stace}},\ }\href {\doibase
  10.1103/physrevb.90.035132} {\bibfield  {journal} {\bibinfo  {journal} {Phys.
  Rev. B}\ }\textbf {\bibinfo {volume} {90}},\ \bibinfo {pages} {035132}
  (\bibinfo {year} {2014})}\BibitemShut {NoStop}%
\bibitem [{\citenamefont {Koshino}\ \emph {et~al.}(2015)\citenamefont
  {Koshino}, \citenamefont {Inomata}, \citenamefont {Lin}, \citenamefont
  {Nakamura},\ and\ \citenamefont {Yamamoto}}]{Koshino2015Theory}%
  \BibitemOpen
  \bibfield  {author} {\bibinfo {author} {\bibfnamefont {K.}~\bibnamefont
  {Koshino}}, \bibinfo {author} {\bibfnamefont {K.}~\bibnamefont {Inomata}},
  \bibinfo {author} {\bibfnamefont {Z.}~\bibnamefont {Lin}}, \bibinfo {author}
  {\bibfnamefont {Y.}~\bibnamefont {Nakamura}}, \ and\ \bibinfo {author}
  {\bibfnamefont {T.}~\bibnamefont {Yamamoto}},\ }\href {\doibase
  10.1103/physreva.91.043805} {\bibfield  {journal} {\bibinfo  {journal} {Phys.
  Rev. A}\ }\textbf {\bibinfo {volume} {91}},\ \bibinfo {pages} {043805}
  (\bibinfo {year} {2015})}\BibitemShut {NoStop}%
\bibitem [{\citenamefont {Inomata}\ \emph {et~al.}()\citenamefont {Inomata},
  \citenamefont {Lin}, \citenamefont {Koshino}, \citenamefont {Oliver},
  \citenamefont {Tsai}, \citenamefont {Yamamoto},\ and\ \citenamefont
  {Nakamura}}]{Inomata2016Single}%
  \BibitemOpen
  \bibfield  {author} {\bibinfo {author} {\bibfnamefont {K.}~\bibnamefont
  {Inomata}}, \bibinfo {author} {\bibfnamefont {Z.}~\bibnamefont {Lin}},
  \bibinfo {author} {\bibfnamefont {K.}~\bibnamefont {Koshino}}, \bibinfo
  {author} {\bibfnamefont {W.~D.}\ \bibnamefont {Oliver}}, \bibinfo {author}
  {\bibfnamefont {J.-S.}\ \bibnamefont {Tsai}}, \bibinfo {author}
  {\bibfnamefont {T.}~\bibnamefont {Yamamoto}}, \ and\ \bibinfo {author}
  {\bibfnamefont {Y.}~\bibnamefont {Nakamura}},\ }\href@noop {} {}\Eprint
  {http://arxiv.org/abs/1601.05513} {arXiv:1601.05513} \BibitemShut {NoStop}%
\bibitem [{\citenamefont {Narla}\ \emph {et~al.}()\citenamefont {Narla},
  \citenamefont {Shankar}, \citenamefont {Hatridge}, \citenamefont {Leghtas},
  \citenamefont {Sliwa}, \citenamefont {Zalys-Geller}, \citenamefont
  {Mundhada}, \citenamefont {Pfaff}, \citenamefont {Frunzio}, \citenamefont
  {Schoelkopf},\ and\ \citenamefont {Devoret}}]{Narla2016Robust}%
  \BibitemOpen
  \bibfield  {author} {\bibinfo {author} {\bibfnamefont {A.}~\bibnamefont
  {Narla}}, \bibinfo {author} {\bibfnamefont {S.}~\bibnamefont {Shankar}},
  \bibinfo {author} {\bibfnamefont {M.}~\bibnamefont {Hatridge}}, \bibinfo
  {author} {\bibfnamefont {Z.}~\bibnamefont {Leghtas}}, \bibinfo {author}
  {\bibfnamefont {K.~M.}\ \bibnamefont {Sliwa}}, \bibinfo {author}
  {\bibfnamefont {E.}~\bibnamefont {Zalys-Geller}}, \bibinfo {author}
  {\bibfnamefont {S.~O.}\ \bibnamefont {Mundhada}}, \bibinfo {author}
  {\bibfnamefont {W.}~\bibnamefont {Pfaff}}, \bibinfo {author} {\bibfnamefont
  {L.}~\bibnamefont {Frunzio}}, \bibinfo {author} {\bibfnamefont {R.~J.}\
  \bibnamefont {Schoelkopf}}, \ and\ \bibinfo {author} {\bibfnamefont {M.~H.}\
  \bibnamefont {Devoret}},\ }\href@noop {} {}\Eprint
  {http://arxiv.org/abs/1603.03742} {arXiv:1603.03742} \BibitemShut {NoStop}%
\bibitem [{\citenamefont {Neumeier}\ \emph {et~al.}(2013)\citenamefont
  {Neumeier}, \citenamefont {Leib},\ and\ \citenamefont
  {Hartmann}}]{Neumeier2013SinglePhoton}%
  \BibitemOpen
  \bibfield  {author} {\bibinfo {author} {\bibfnamefont {L.}~\bibnamefont
  {Neumeier}}, \bibinfo {author} {\bibfnamefont {M.}~\bibnamefont {Leib}}, \
  and\ \bibinfo {author} {\bibfnamefont {M.~J.}\ \bibnamefont {Hartmann}},\
  }\href {\doibase 10.1103/physrevlett.111.063601} {\bibfield  {journal}
  {\bibinfo  {journal} {Phys. Rev. Lett.}\ }\textbf {\bibinfo {volume} {111}},\
  \bibinfo {pages} {063601} (\bibinfo {year} {2013})}\BibitemShut {NoStop}%
\bibitem [{\citenamefont {Manzoni}\ \emph {et~al.}(2014)\citenamefont
  {Manzoni}, \citenamefont {Reiter}, \citenamefont {Taylor},\ and\
  \citenamefont {S{\o}{}rensen}}]{Manzoni2014Singlephoton}%
  \BibitemOpen
  \bibfield  {author} {\bibinfo {author} {\bibfnamefont {M.~T.}\ \bibnamefont
  {Manzoni}}, \bibinfo {author} {\bibfnamefont {F.}~\bibnamefont {Reiter}},
  \bibinfo {author} {\bibfnamefont {J.~M.}\ \bibnamefont {Taylor}}, \ and\
  \bibinfo {author} {\bibfnamefont {A.~S.}\ \bibnamefont {S{\o}{}rensen}},\
  }\href {\doibase 10.1103/physrevb.89.180502} {\bibfield  {journal} {\bibinfo
  {journal} {Phys. Rev. B}\ }\textbf {\bibinfo {volume} {89}},\ \bibinfo
  {pages} {180502} (\bibinfo {year} {2014})}\BibitemShut {NoStop}%
\bibitem [{\citenamefont {Cirac}\ \emph {et~al.}(1997)\citenamefont {Cirac},
  \citenamefont {Zoller}, \citenamefont {Kimble},\ and\ \citenamefont
  {Mabuchi}}]{Cirac1997Quantum}%
  \BibitemOpen
  \bibfield  {author} {\bibinfo {author} {\bibfnamefont {J.~I.}\ \bibnamefont
  {Cirac}}, \bibinfo {author} {\bibfnamefont {P.}~\bibnamefont {Zoller}},
  \bibinfo {author} {\bibfnamefont {H.~J.}\ \bibnamefont {Kimble}}, \ and\
  \bibinfo {author} {\bibfnamefont {H.}~\bibnamefont {Mabuchi}},\ }\href
  {\doibase 10.1103/physrevlett.78.3221} {\bibfield  {journal} {\bibinfo
  {journal} {Phys. Rev. Lett.}\ }\textbf {\bibinfo {volume} {78}},\ \bibinfo
  {pages} {3221} (\bibinfo {year} {1997})}\BibitemShut {NoStop}%
\bibitem [{\citenamefont {Yin}\ \emph {et~al.}(2013)\citenamefont {Yin},
  \citenamefont {Chen}, \citenamefont {Sank}, \citenamefont {O'Malley},
  \citenamefont {White}, \citenamefont {Barends}, \citenamefont {Kelly},
  \citenamefont {Lucero}, \citenamefont {Mariantoni}, \citenamefont {Megrant},
  \citenamefont {Neill}, \citenamefont {Vainsencher}, \citenamefont {Wenner},
  \citenamefont {Korotkov}, \citenamefont {Cleland},\ and\ \citenamefont
  {Martinis}}]{Yin2013Catch}%
  \BibitemOpen
  \bibfield  {author} {\bibinfo {author} {\bibfnamefont {Y.}~\bibnamefont
  {Yin}}, \bibinfo {author} {\bibfnamefont {Y.}~\bibnamefont {Chen}}, \bibinfo
  {author} {\bibfnamefont {D.}~\bibnamefont {Sank}}, \bibinfo {author}
  {\bibfnamefont {P.~J.~J.}\ \bibnamefont {O'Malley}}, \bibinfo {author}
  {\bibfnamefont {T.~C.}\ \bibnamefont {White}}, \bibinfo {author}
  {\bibfnamefont {R.}~\bibnamefont {Barends}}, \bibinfo {author} {\bibfnamefont
  {J.}~\bibnamefont {Kelly}}, \bibinfo {author} {\bibfnamefont
  {E.}~\bibnamefont {Lucero}}, \bibinfo {author} {\bibfnamefont
  {M.}~\bibnamefont {Mariantoni}}, \bibinfo {author} {\bibfnamefont
  {A.}~\bibnamefont {Megrant}}, \bibinfo {author} {\bibfnamefont
  {C.}~\bibnamefont {Neill}}, \bibinfo {author} {\bibfnamefont
  {A.}~\bibnamefont {Vainsencher}}, \bibinfo {author} {\bibfnamefont
  {J.}~\bibnamefont {Wenner}}, \bibinfo {author} {\bibfnamefont {A.~N.}\
  \bibnamefont {Korotkov}}, \bibinfo {author} {\bibfnamefont {A.~N.}\
  \bibnamefont {Cleland}}, \ and\ \bibinfo {author} {\bibfnamefont {J.~M.}\
  \bibnamefont {Martinis}},\ }\href {\doibase 10.1103/physrevlett.110.107001}
  {\bibfield  {journal} {\bibinfo  {journal} {Phys. Rev. Lett.}\ }\textbf
  {\bibinfo {volume} {110}},\ \bibinfo {pages} {107001} (\bibinfo {year}
  {2013})}\BibitemShut {NoStop}%
\bibitem [{\citenamefont {Srinivasan}\ \emph {et~al.}(2014)\citenamefont
  {Srinivasan}, \citenamefont {Sundaresan}, \citenamefont {Sadri},
  \citenamefont {Liu}, \citenamefont {Gambetta}, \citenamefont {Yu},
  \citenamefont {Girvin},\ and\ \citenamefont
  {Houck}}]{Srinivasan2014Timereversal}%
  \BibitemOpen
  \bibfield  {author} {\bibinfo {author} {\bibfnamefont {S.~J.}\ \bibnamefont
  {Srinivasan}}, \bibinfo {author} {\bibfnamefont {N.~M.}\ \bibnamefont
  {Sundaresan}}, \bibinfo {author} {\bibfnamefont {D.}~\bibnamefont {Sadri}},
  \bibinfo {author} {\bibfnamefont {Y.}~\bibnamefont {Liu}}, \bibinfo {author}
  {\bibfnamefont {J.~M.}\ \bibnamefont {Gambetta}}, \bibinfo {author}
  {\bibfnamefont {T.}~\bibnamefont {Yu}}, \bibinfo {author} {\bibfnamefont
  {S.~M.}\ \bibnamefont {Girvin}}, \ and\ \bibinfo {author} {\bibfnamefont
  {A.~A.}\ \bibnamefont {Houck}},\ }\href {\doibase 10.1103/physreva.89.033857}
  {\bibfield  {journal} {\bibinfo  {journal} {Phys. Rev. A}\ }\textbf {\bibinfo
  {volume} {89}},\ \bibinfo {pages} {033857} (\bibinfo {year}
  {2014})}\BibitemShut {NoStop}%
\bibitem [{\citenamefont {Wenner}\ \emph {et~al.}(2014)\citenamefont {Wenner},
  \citenamefont {Yin}, \citenamefont {Chen}, \citenamefont {Barends},
  \citenamefont {Chiaro}, \citenamefont {Jeffrey}, \citenamefont {Kelly},
  \citenamefont {Megrant}, \citenamefont {Mutus}, \citenamefont {Neill},
  \citenamefont {O'Malley}, \citenamefont {Roushan}, \citenamefont {Sank},
  \citenamefont {Vainsencher}, \citenamefont {White}, \citenamefont {Korotkov},
  \citenamefont {Cleland},\ and\ \citenamefont
  {Martinis}}]{Wenner2014Catching}%
  \BibitemOpen
  \bibfield  {author} {\bibinfo {author} {\bibfnamefont {J.}~\bibnamefont
  {Wenner}}, \bibinfo {author} {\bibfnamefont {Y.}~\bibnamefont {Yin}},
  \bibinfo {author} {\bibfnamefont {Y.}~\bibnamefont {Chen}}, \bibinfo {author}
  {\bibfnamefont {R.}~\bibnamefont {Barends}}, \bibinfo {author} {\bibfnamefont
  {B.}~\bibnamefont {Chiaro}}, \bibinfo {author} {\bibfnamefont
  {E.}~\bibnamefont {Jeffrey}}, \bibinfo {author} {\bibfnamefont
  {J.}~\bibnamefont {Kelly}}, \bibinfo {author} {\bibfnamefont
  {A.}~\bibnamefont {Megrant}}, \bibinfo {author} {\bibfnamefont {J.~Y.}\
  \bibnamefont {Mutus}}, \bibinfo {author} {\bibfnamefont {C.}~\bibnamefont
  {Neill}}, \bibinfo {author} {\bibfnamefont {P.~J.~J.}\ \bibnamefont
  {O'Malley}}, \bibinfo {author} {\bibfnamefont {P.}~\bibnamefont {Roushan}},
  \bibinfo {author} {\bibfnamefont {D.}~\bibnamefont {Sank}}, \bibinfo {author}
  {\bibfnamefont {A.}~\bibnamefont {Vainsencher}}, \bibinfo {author}
  {\bibfnamefont {T.~C.}\ \bibnamefont {White}}, \bibinfo {author}
  {\bibfnamefont {A.~N.}\ \bibnamefont {Korotkov}}, \bibinfo {author}
  {\bibfnamefont {A.~N.}\ \bibnamefont {Cleland}}, \ and\ \bibinfo {author}
  {\bibfnamefont {J.~M.}\ \bibnamefont {Martinis}},\ }\href {\doibase
  10.1103/physrevlett.112.210501} {\bibfield  {journal} {\bibinfo  {journal}
  {Phys. Rev. Lett.}\ }\textbf {\bibinfo {volume} {112}},\ \bibinfo {pages}
  {210501} (\bibinfo {year} {2014})}\BibitemShut {NoStop}%
\bibitem [{\citenamefont {Pechal}\ \emph {et~al.}(2014)\citenamefont {Pechal},
  \citenamefont {Huthmacher}, \citenamefont {Eichler}, \citenamefont
  {Zeytino\u{g}lu}, \citenamefont {Abdumalikov}, \citenamefont {Berger},
  \citenamefont {Wallraff},\ and\ \citenamefont
  {Filipp}}]{Pechal2014MicrowaveControlled}%
  \BibitemOpen
  \bibfield  {author} {\bibinfo {author} {\bibfnamefont {M.}~\bibnamefont
  {Pechal}}, \bibinfo {author} {\bibfnamefont {L.}~\bibnamefont {Huthmacher}},
  \bibinfo {author} {\bibfnamefont {C.}~\bibnamefont {Eichler}}, \bibinfo
  {author} {\bibfnamefont {S.}~\bibnamefont {Zeytino\u{g}lu}}, \bibinfo
  {author} {\bibfnamefont {A.~A.}\ \bibnamefont {Abdumalikov}}, \bibinfo
  {author} {\bibfnamefont {S.}~\bibnamefont {Berger}}, \bibinfo {author}
  {\bibfnamefont {A.}~\bibnamefont {Wallraff}}, \ and\ \bibinfo {author}
  {\bibfnamefont {S.}~\bibnamefont {Filipp}},\ }\href {\doibase
  10.1103/physrevx.4.041010} {\bibfield  {journal} {\bibinfo  {journal} {Phys.
  Rev. X}\ }\textbf {\bibinfo {volume} {4}},\ \bibinfo {pages} {041010}
  (\bibinfo {year} {2014})}\BibitemShut {NoStop}%
\bibitem [{\citenamefont {Gleyzes}\ \emph {et~al.}(2007)\citenamefont
  {Gleyzes}, \citenamefont {Kuhr}, \citenamefont {Guerlin}, \citenamefont
  {Bernu}, \citenamefont {Del\'{e}glise}, \citenamefont {Busk~Hoff},
  \citenamefont {Brune}, \citenamefont {Raimond},\ and\ \citenamefont
  {Haroche}}]{Gleyzes2007Quantum}%
  \BibitemOpen
  \bibfield  {author} {\bibinfo {author} {\bibfnamefont {S.}~\bibnamefont
  {Gleyzes}}, \bibinfo {author} {\bibfnamefont {S.}~\bibnamefont {Kuhr}},
  \bibinfo {author} {\bibfnamefont {C.}~\bibnamefont {Guerlin}}, \bibinfo
  {author} {\bibfnamefont {J.}~\bibnamefont {Bernu}}, \bibinfo {author}
  {\bibfnamefont {S.}~\bibnamefont {Del\'{e}glise}}, \bibinfo {author}
  {\bibfnamefont {U.}~\bibnamefont {Busk~Hoff}}, \bibinfo {author}
  {\bibfnamefont {M.}~\bibnamefont {Brune}}, \bibinfo {author} {\bibfnamefont
  {J.-M.}\ \bibnamefont {Raimond}}, \ and\ \bibinfo {author} {\bibfnamefont
  {S.}~\bibnamefont {Haroche}},\ }\href {\doibase 10.1038/nature05589}
  {\bibfield  {journal} {\bibinfo  {journal} {Nature (London)}\ }\textbf
  {\bibinfo {volume} {446}},\ \bibinfo {pages} {297} (\bibinfo {year}
  {2007})}\BibitemShut {NoStop}%
\bibitem [{\citenamefont {Sun}\ \emph {et~al.}(2014)\citenamefont {Sun},
  \citenamefont {Petrenko}, \citenamefont {Leghtas}, \citenamefont {Vlastakis},
  \citenamefont {Kirchmair}, \citenamefont {Sliwa}, \citenamefont {Narla},
  \citenamefont {Hatridge}, \citenamefont {Shankar}, \citenamefont {Blumoff},
  \citenamefont {Frunzio}, \citenamefont {Mirrahimi}, \citenamefont {Devoret},\
  and\ \citenamefont {Schoelkopf}}]{Sun2014Tracking}%
  \BibitemOpen
  \bibfield  {author} {\bibinfo {author} {\bibfnamefont {L.}~\bibnamefont
  {Sun}}, \bibinfo {author} {\bibfnamefont {A.}~\bibnamefont {Petrenko}},
  \bibinfo {author} {\bibfnamefont {Z.}~\bibnamefont {Leghtas}}, \bibinfo
  {author} {\bibfnamefont {B.}~\bibnamefont {Vlastakis}}, \bibinfo {author}
  {\bibfnamefont {G.}~\bibnamefont {Kirchmair}}, \bibinfo {author}
  {\bibfnamefont {K.~M.}\ \bibnamefont {Sliwa}}, \bibinfo {author}
  {\bibfnamefont {A.}~\bibnamefont {Narla}}, \bibinfo {author} {\bibfnamefont
  {M.}~\bibnamefont {Hatridge}}, \bibinfo {author} {\bibfnamefont
  {S.}~\bibnamefont {Shankar}}, \bibinfo {author} {\bibfnamefont
  {J.}~\bibnamefont {Blumoff}}, \bibinfo {author} {\bibfnamefont
  {L.}~\bibnamefont {Frunzio}}, \bibinfo {author} {\bibfnamefont
  {M.}~\bibnamefont {Mirrahimi}}, \bibinfo {author} {\bibfnamefont {M.~H.}\
  \bibnamefont {Devoret}}, \ and\ \bibinfo {author} {\bibfnamefont {R.~J.}\
  \bibnamefont {Schoelkopf}},\ }\href {\doibase 10.1038/nature13436} {\bibfield
   {journal} {\bibinfo  {journal} {Nature (London)}\ }\textbf {\bibinfo
  {volume} {511}},\ \bibinfo {pages} {444} (\bibinfo {year}
  {2014})}\BibitemShut {NoStop}%
\bibitem [{Note1()}]{Note1}%
  \BibitemOpen
  \bibinfo {note} {See the Supplemental Material for additional single-shot
  histograms and for details of the experimental setup, the circuit model, the
  numerical model, and the measurements of $\protect \mathaccentV
  {tilde}07E{G}$, $\tau $, and $C_{e}$.}\BibitemShut {Stop}%
\bibitem [{\citenamefont {Chiorescu}\ \emph {et~al.}(2003)\citenamefont
  {Chiorescu}, \citenamefont {Nakamura}, \citenamefont {Harmans},\ and\
  \citenamefont {Mooij}}]{Chiorescu2003Coherent}%
  \BibitemOpen
  \bibfield  {author} {\bibinfo {author} {\bibfnamefont {I.}~\bibnamefont
  {Chiorescu}}, \bibinfo {author} {\bibfnamefont {Y.}~\bibnamefont {Nakamura}},
  \bibinfo {author} {\bibfnamefont {C.~J. P.~M.}\ \bibnamefont {Harmans}}, \
  and\ \bibinfo {author} {\bibfnamefont {J.~E.}\ \bibnamefont {Mooij}},\ }\href
  {\doibase 10.1126/science.1081045} {\bibfield  {journal} {\bibinfo  {journal}
  {Science}\ }\textbf {\bibinfo {volume} {299}},\ \bibinfo {pages} {1869}
  (\bibinfo {year} {2003})}\BibitemShut {NoStop}%
\bibitem [{\citenamefont {Govenius}\ \emph {et~al.}(2014)\citenamefont
  {Govenius}, \citenamefont {Lake}, \citenamefont {Tan}, \citenamefont
  {Pietil\"{a}}, \citenamefont {Julin}, \citenamefont {Maasilta}, \citenamefont
  {Virtanen},\ and\ \citenamefont {M\"{o}tt\"{o}nen}}]{Govenius2014Microwave}%
  \BibitemOpen
  \bibfield  {author} {\bibinfo {author} {\bibfnamefont {J.}~\bibnamefont
  {Govenius}}, \bibinfo {author} {\bibfnamefont {R.~E.}\ \bibnamefont {Lake}},
  \bibinfo {author} {\bibfnamefont {K.~Y.}\ \bibnamefont {Tan}}, \bibinfo
  {author} {\bibfnamefont {V.}~\bibnamefont {Pietil\"{a}}}, \bibinfo {author}
  {\bibfnamefont {J.~K.}\ \bibnamefont {Julin}}, \bibinfo {author}
  {\bibfnamefont {I.~J.}\ \bibnamefont {Maasilta}}, \bibinfo {author}
  {\bibfnamefont {P.}~\bibnamefont {Virtanen}}, \ and\ \bibinfo {author}
  {\bibfnamefont {M.}~\bibnamefont {M\"{o}tt\"{o}nen}},\ }\href {\doibase
  10.1103/physrevb.90.064505} {\bibfield  {journal} {\bibinfo  {journal} {Phys.
  Rev. B}\ }\textbf {\bibinfo {volume} {90}},\ \bibinfo {pages} {064505}
  (\bibinfo {year} {2014})}\BibitemShut {NoStop}%
\bibitem [{\citenamefont {Santavicca}\ \emph {et~al.}(2010)\citenamefont
  {Santavicca}, \citenamefont {Reulet}, \citenamefont {Karasik}, \citenamefont
  {Pereverzev}, \citenamefont {Olaya}, \citenamefont {Gershenson},
  \citenamefont {Frunzio},\ and\ \citenamefont
  {Prober}}]{Santavicca2010Energy}%
  \BibitemOpen
  \bibfield  {author} {\bibinfo {author} {\bibfnamefont {D.~F.}\ \bibnamefont
  {Santavicca}}, \bibinfo {author} {\bibfnamefont {B.}~\bibnamefont {Reulet}},
  \bibinfo {author} {\bibfnamefont {B.~S.}\ \bibnamefont {Karasik}}, \bibinfo
  {author} {\bibfnamefont {S.~V.}\ \bibnamefont {Pereverzev}}, \bibinfo
  {author} {\bibfnamefont {D.}~\bibnamefont {Olaya}}, \bibinfo {author}
  {\bibfnamefont {M.~E.}\ \bibnamefont {Gershenson}}, \bibinfo {author}
  {\bibfnamefont {L.}~\bibnamefont {Frunzio}}, \ and\ \bibinfo {author}
  {\bibfnamefont {D.~E.}\ \bibnamefont {Prober}},\ }\href {\doibase
  10.1063/1.3336008} {\bibfield  {journal} {\bibinfo  {journal} {Appl. Phys.
  Lett.}\ }\textbf {\bibinfo {volume} {96}},\ \bibinfo {pages} {083505}
  (\bibinfo {year} {2010})}\BibitemShut {NoStop}%
\bibitem [{\citenamefont {de~Visser}\ \emph {et~al.}(2010)\citenamefont
  {de~Visser}, \citenamefont {Withington},\ and\ \citenamefont
  {Goldie}}]{deVisser2010Readoutpower}%
  \BibitemOpen
  \bibfield  {author} {\bibinfo {author} {\bibfnamefont {P.~J.}\ \bibnamefont
  {de~Visser}}, \bibinfo {author} {\bibfnamefont {S.}~\bibnamefont
  {Withington}}, \ and\ \bibinfo {author} {\bibfnamefont {D.~J.}\ \bibnamefont
  {Goldie}},\ }\href {\doibase 10.1063/1.3517152} {\bibfield  {journal}
  {\bibinfo  {journal} {J. Appl. Phys.}\ }\textbf {\bibinfo {volume} {108}},\
  \bibinfo {pages} {114504} (\bibinfo {year} {2010})}\BibitemShut {NoStop}%
\bibitem [{\citenamefont {Thompson}\ \emph {et~al.}(2013)\citenamefont
  {Thompson}, \citenamefont {Withington}, \citenamefont {Goldie},\ and\
  \citenamefont {Thomas}}]{Thompson2013Dynamical}%
  \BibitemOpen
  \bibfield  {author} {\bibinfo {author} {\bibfnamefont {S.~E.}\ \bibnamefont
  {Thompson}}, \bibinfo {author} {\bibfnamefont {S.}~\bibnamefont
  {Withington}}, \bibinfo {author} {\bibfnamefont {D.~J.}\ \bibnamefont
  {Goldie}}, \ and\ \bibinfo {author} {\bibfnamefont {C.~N.}\ \bibnamefont
  {Thomas}},\ }\href {\doibase 10.1088/0953-2048/26/9/095009} {\bibfield
  {journal} {\bibinfo  {journal} {Supercond. Sci. Technol.}\ }\textbf {\bibinfo
  {volume} {26}},\ \bibinfo {pages} {095009} (\bibinfo {year}
  {2013})}\BibitemShut {NoStop}%
\bibitem [{\citenamefont {Lindeman}(2014)}]{Lindeman2014Resonatorbolometer}%
  \BibitemOpen
  \bibfield  {author} {\bibinfo {author} {\bibfnamefont {M.~A.}\ \bibnamefont
  {Lindeman}},\ }\href {\doibase 10.1063/1.4890018} {\bibfield  {journal}
  {\bibinfo  {journal} {J. Appl. Phys.}\ }\textbf {\bibinfo {volume} {116}},\
  \bibinfo {pages} {024506} (\bibinfo {year} {2014})}\BibitemShut {NoStop}%
\bibitem [{\citenamefont {Thomas}\ \emph {et~al.}(2015)\citenamefont {Thomas},
  \citenamefont {Withington},\ and\ \citenamefont
  {Goldie}}]{Thomas2015Electrothermal}%
  \BibitemOpen
  \bibfield  {author} {\bibinfo {author} {\bibfnamefont {C.~N.}\ \bibnamefont
  {Thomas}}, \bibinfo {author} {\bibfnamefont {S.}~\bibnamefont {Withington}},
  \ and\ \bibinfo {author} {\bibfnamefont {D.~J.}\ \bibnamefont {Goldie}},\
  }\href {\doibase 10.1088/0953-2048/28/4/045012} {\bibfield  {journal}
  {\bibinfo  {journal} {Supercond. Sci. Technol.}\ }\textbf {\bibinfo {volume}
  {28}},\ \bibinfo {pages} {045012} (\bibinfo {year} {2015})}\BibitemShut
  {NoStop}%
\bibitem [{\citenamefont {Braginsky}\ \emph {et~al.}(1989)\citenamefont
  {Braginsky}, \citenamefont {Gorodetsky},\ and\ \citenamefont
  {Ilchenko}}]{Braginsky1989Qualityfactor}%
  \BibitemOpen
  \bibfield  {author} {\bibinfo {author} {\bibfnamefont {V.~B.}\ \bibnamefont
  {Braginsky}}, \bibinfo {author} {\bibfnamefont {M.~L.}\ \bibnamefont
  {Gorodetsky}}, \ and\ \bibinfo {author} {\bibfnamefont {V.~S.}\ \bibnamefont
  {Ilchenko}},\ }\href {\doibase 10.1016/0375-9601(89)90912-2} {\bibfield
  {journal} {\bibinfo  {journal} {Phys. Lett. A}\ }\textbf {\bibinfo {volume}
  {137}},\ \bibinfo {pages} {393} (\bibinfo {year} {1989})}\BibitemShut
  {NoStop}%
\bibitem [{\citenamefont {Fomin}\ \emph {et~al.}(2005)\citenamefont {Fomin},
  \citenamefont {Gorodetsky}, \citenamefont {Grudinin},\ and\ \citenamefont
  {Ilchenko}}]{Fomin2005Nonstationary}%
  \BibitemOpen
  \bibfield  {author} {\bibinfo {author} {\bibfnamefont {A.~E.}\ \bibnamefont
  {Fomin}}, \bibinfo {author} {\bibfnamefont {M.~L.}\ \bibnamefont
  {Gorodetsky}}, \bibinfo {author} {\bibfnamefont {I.~S.}\ \bibnamefont
  {Grudinin}}, \ and\ \bibinfo {author} {\bibfnamefont {V.~S.}\ \bibnamefont
  {Ilchenko}},\ }\href {\doibase 10.1364/josab.22.000459} {\bibfield  {journal}
  {\bibinfo  {journal} {J. Opt. Soc. Am. B}\ }\textbf {\bibinfo {volume}
  {22}},\ \bibinfo {pages} {459} (\bibinfo {year} {2005})}\BibitemShut
  {NoStop}%
\bibitem [{\citenamefont {Irwin}(1995)}]{Irwin1995Application}%
  \BibitemOpen
  \bibfield  {author} {\bibinfo {author} {\bibfnamefont {K.~D.}\ \bibnamefont
  {Irwin}},\ }\href {\doibase 10.1063/1.113674} {\bibfield  {journal} {\bibinfo
   {journal} {Appl. Phys. Lett.}\ }\textbf {\bibinfo {volume} {66}},\ \bibinfo
  {pages} {1998} (\bibinfo {year} {1995})}\BibitemShut {NoStop}%
\bibitem [{\citenamefont {Nahum}\ and\ \citenamefont
  {Martinis}(1993)}]{Nahum1993Ultrasensitivehotelectron}%
  \BibitemOpen
  \bibfield  {author} {\bibinfo {author} {\bibfnamefont {M.}~\bibnamefont
  {Nahum}}\ and\ \bibinfo {author} {\bibfnamefont {J.~M.}\ \bibnamefont
  {Martinis}},\ }\href {\doibase 10.1063/1.110237} {\bibfield  {journal}
  {\bibinfo  {journal} {Appl. Phys. Lett.}\ }\textbf {\bibinfo {volume} {63}},\
  \bibinfo {pages} {3075} (\bibinfo {year} {1993})}\BibitemShut {NoStop}%
\bibitem [{\citenamefont {Giazotto}\ \emph {et~al.}(2008)\citenamefont
  {Giazotto}, \citenamefont {Heikkil\"{a}}, \citenamefont {Pepe}, \citenamefont
  {Helist\"{o}}, \citenamefont {Luukanen},\ and\ \citenamefont
  {Pekola}}]{Giazotto2008Ultrasensitive}%
  \BibitemOpen
  \bibfield  {author} {\bibinfo {author} {\bibfnamefont {F.}~\bibnamefont
  {Giazotto}}, \bibinfo {author} {\bibfnamefont {T.~T.}\ \bibnamefont
  {Heikkil\"{a}}}, \bibinfo {author} {\bibfnamefont {G.~P.}\ \bibnamefont
  {Pepe}}, \bibinfo {author} {\bibfnamefont {P.}~\bibnamefont {Helist\"{o}}},
  \bibinfo {author} {\bibfnamefont {A.}~\bibnamefont {Luukanen}}, \ and\
  \bibinfo {author} {\bibfnamefont {J.~P.}\ \bibnamefont {Pekola}},\ }\href
  {\doibase 10.1063/1.2908922} {\bibfield  {journal} {\bibinfo  {journal}
  {Appl. Phys. Lett.}\ }\textbf {\bibinfo {volume} {92}},\ \bibinfo {pages}
  {162507} (\bibinfo {year} {2008})}\BibitemShut {NoStop}%
\bibitem [{\citenamefont {Voutilainen}\ \emph {et~al.}(2010)\citenamefont
  {Voutilainen}, \citenamefont {Laakso},\ and\ \citenamefont
  {Heikkil\"{a}}}]{Voutilainen2010Physics}%
  \BibitemOpen
  \bibfield  {author} {\bibinfo {author} {\bibfnamefont {J.}~\bibnamefont
  {Voutilainen}}, \bibinfo {author} {\bibfnamefont {M.~A.}\ \bibnamefont
  {Laakso}}, \ and\ \bibinfo {author} {\bibfnamefont {T.~T.}\ \bibnamefont
  {Heikkil\"{a}}},\ }\href {\doibase 10.1063/1.3354042} {\bibfield  {journal}
  {\bibinfo  {journal} {J. Appl. Phys.}\ }\textbf {\bibinfo {volume} {107}},\
  \bibinfo {pages} {064508} (\bibinfo {year} {2010})}\BibitemShut {NoStop}%
\bibitem [{\citenamefont {Timofeev}\ \emph {et~al.}(2009)\citenamefont
  {Timofeev}, \citenamefont {Helle}, \citenamefont {Meschke}, \citenamefont
  {M\"{o}tt\"{o}nen},\ and\ \citenamefont {Pekola}}]{Timofeev2009Electronic}%
  \BibitemOpen
  \bibfield  {author} {\bibinfo {author} {\bibfnamefont {A.~V.}\ \bibnamefont
  {Timofeev}}, \bibinfo {author} {\bibfnamefont {M.}~\bibnamefont {Helle}},
  \bibinfo {author} {\bibfnamefont {M.}~\bibnamefont {Meschke}}, \bibinfo
  {author} {\bibfnamefont {M.}~\bibnamefont {M\"{o}tt\"{o}nen}}, \ and\
  \bibinfo {author} {\bibfnamefont {J.~P.}\ \bibnamefont {Pekola}},\ }\href
  {\doibase 10.1103/physrevlett.102.200801} {\bibfield  {journal} {\bibinfo
  {journal} {Phys. Rev. Lett.}\ }\textbf {\bibinfo {volume} {102}},\ \bibinfo
  {pages} {200801} (\bibinfo {year} {2009})}\BibitemShut {NoStop}%
\bibitem [{\citenamefont {Pendry}(1983)}]{Pendry1983Quantum}%
  \BibitemOpen
  \bibfield  {author} {\bibinfo {author} {\bibfnamefont {J.~B.}\ \bibnamefont
  {Pendry}},\ }\href {\doibase 10.1088/0305-4470/16/10/012} {\bibfield
  {journal} {\bibinfo  {journal} {J. Phys. A: Math. Gen.}\ }\textbf {\bibinfo
  {volume} {16}},\ \bibinfo {pages} {2161} (\bibinfo {year}
  {1983})}\BibitemShut {NoStop}%
\bibitem [{\citenamefont {Schmidt}\ \emph {et~al.}(2004)\citenamefont
  {Schmidt}, \citenamefont {Schoelkopf},\ and\ \citenamefont
  {Cleland}}]{Schmidt2004PhotonMediated}%
  \BibitemOpen
  \bibfield  {author} {\bibinfo {author} {\bibfnamefont {D.~R.}\ \bibnamefont
  {Schmidt}}, \bibinfo {author} {\bibfnamefont {R.~J.}\ \bibnamefont
  {Schoelkopf}}, \ and\ \bibinfo {author} {\bibfnamefont {A.~N.}\ \bibnamefont
  {Cleland}},\ }\href {\doibase 10.1103/physrevlett.93.045901} {\bibfield
  {journal} {\bibinfo  {journal} {Phys. Rev. Lett.}\ }\textbf {\bibinfo
  {volume} {93}},\ \bibinfo {pages} {045901} (\bibinfo {year}
  {2004})}\BibitemShut {NoStop}%
\bibitem [{\citenamefont {Meschke}\ \emph {et~al.}(2006)\citenamefont
  {Meschke}, \citenamefont {Guichard},\ and\ \citenamefont
  {Pekola}}]{Meschke2006Singlemode}%
  \BibitemOpen
  \bibfield  {author} {\bibinfo {author} {\bibfnamefont {M.}~\bibnamefont
  {Meschke}}, \bibinfo {author} {\bibfnamefont {W.}~\bibnamefont {Guichard}}, \
  and\ \bibinfo {author} {\bibfnamefont {J.~P.}\ \bibnamefont {Pekola}},\
  }\href {\doibase 10.1038/nature05276} {\bibfield  {journal} {\bibinfo
  {journal} {Nature (London)}\ }\textbf {\bibinfo {volume} {444}},\ \bibinfo
  {pages} {187} (\bibinfo {year} {2006})}\BibitemShut {NoStop}%
\bibitem [{\citenamefont {Partanen}\ \emph {et~al.}(2016)\citenamefont
  {Partanen}, \citenamefont {Tan}, \citenamefont {Govenius}, \citenamefont
  {Lake}, \citenamefont {M\"{a}kel\"{a}}, \citenamefont {Tanttu},\ and\
  \citenamefont {M\"{o}tt\"{o}nen}}]{Partanen2016Quantumlimited}%
  \BibitemOpen
  \bibfield  {author} {\bibinfo {author} {\bibfnamefont {M.}~\bibnamefont
  {Partanen}}, \bibinfo {author} {\bibfnamefont {K.~Y.}\ \bibnamefont {Tan}},
  \bibinfo {author} {\bibfnamefont {J.}~\bibnamefont {Govenius}}, \bibinfo
  {author} {\bibfnamefont {R.~E.}\ \bibnamefont {Lake}}, \bibinfo {author}
  {\bibfnamefont {M.~K.}\ \bibnamefont {M\"{a}kel\"{a}}}, \bibinfo {author}
  {\bibfnamefont {T.}~\bibnamefont {Tanttu}}, \ and\ \bibinfo {author}
  {\bibfnamefont {M.}~\bibnamefont {M\"{o}tt\"{o}nen}},\ }\href {\doibase
  10.1038/nphys3642} {\bibfield  {journal} {\bibinfo  {journal} {Nature Phys.}\
  }\textbf {\bibinfo {volume} {12}},\ \bibinfo {pages} {460} (\bibinfo {year}
  {2016})}\BibitemShut {NoStop}%
\bibitem [{\citenamefont {Zimmer}(1967)}]{Zimmer1967PARAMETRIC}%
  \BibitemOpen
  \bibfield  {author} {\bibinfo {author} {\bibfnamefont {H.}~\bibnamefont
  {Zimmer}},\ }\href {\doibase 10.1063/1.1754906} {\bibfield  {journal}
  {\bibinfo  {journal} {Appl. Phys. Lett.}\ }\textbf {\bibinfo {volume} {10}},\
  \bibinfo {pages} {193} (\bibinfo {year} {1967})}\BibitemShut {NoStop}%
\bibitem [{\citenamefont {Castellanos-Beltran}\ and\ \citenamefont
  {Lehnert}(2007)}]{CastellanosBeltran2007Widely}%
  \BibitemOpen
  \bibfield  {author} {\bibinfo {author} {\bibfnamefont {M.~A.}\ \bibnamefont
  {Castellanos-Beltran}}\ and\ \bibinfo {author} {\bibfnamefont {K.~W.}\
  \bibnamefont {Lehnert}},\ }\href {\doibase 10.1063/1.2773988} {\bibfield
  {journal} {\bibinfo  {journal} {Appl. Phys. Lett.}\ }\textbf {\bibinfo
  {volume} {91}},\ \bibinfo {pages} {083509} (\bibinfo {year}
  {2007})}\BibitemShut {NoStop}%
\bibitem [{\citenamefont {Siddiqi}\ \emph {et~al.}(2004)\citenamefont
  {Siddiqi}, \citenamefont {Vijay}, \citenamefont {Pierre}, \citenamefont
  {Wilson}, \citenamefont {Metcalfe}, \citenamefont {Rigetti}, \citenamefont
  {Frunzio},\ and\ \citenamefont {Devoret}}]{Siddiqi2004RFDriven}%
  \BibitemOpen
  \bibfield  {author} {\bibinfo {author} {\bibfnamefont {I.}~\bibnamefont
  {Siddiqi}}, \bibinfo {author} {\bibfnamefont {R.}~\bibnamefont {Vijay}},
  \bibinfo {author} {\bibfnamefont {F.}~\bibnamefont {Pierre}}, \bibinfo
  {author} {\bibfnamefont {C.~M.}\ \bibnamefont {Wilson}}, \bibinfo {author}
  {\bibfnamefont {M.}~\bibnamefont {Metcalfe}}, \bibinfo {author}
  {\bibfnamefont {C.}~\bibnamefont {Rigetti}}, \bibinfo {author} {\bibfnamefont
  {L.}~\bibnamefont {Frunzio}}, \ and\ \bibinfo {author} {\bibfnamefont
  {M.~H.}\ \bibnamefont {Devoret}},\ }\href {\doibase
  10.1103/physrevlett.93.207002} {\bibfield  {journal} {\bibinfo  {journal}
  {Phys. Rev. Lett.}\ }\textbf {\bibinfo {volume} {93}},\ \bibinfo {pages}
  {207002} (\bibinfo {year} {2004})}\BibitemShut {NoStop}%
\bibitem [{\citenamefont {Vijay}\ \emph {et~al.}(2009)\citenamefont {Vijay},
  \citenamefont {Devoret},\ and\ \citenamefont {Siddiqi}}]{Vijay2009Invited}%
  \BibitemOpen
  \bibfield  {author} {\bibinfo {author} {\bibfnamefont {R.}~\bibnamefont
  {Vijay}}, \bibinfo {author} {\bibfnamefont {M.~H.}\ \bibnamefont {Devoret}},
  \ and\ \bibinfo {author} {\bibfnamefont {I.}~\bibnamefont {Siddiqi}},\ }\href
  {\doibase 10.1063/1.3224703} {\bibfield  {journal} {\bibinfo  {journal} {Rev.
  Sci. Instrum.}\ }\textbf {\bibinfo {volume} {80}},\ \bibinfo {pages} {111101}
  (\bibinfo {year} {2009})}\BibitemShut {NoStop}%
\bibitem [{\citenamefont {Moseley}\ \emph {et~al.}(1984)\citenamefont
  {Moseley}, \citenamefont {Mather},\ and\ \citenamefont
  {McCammon}}]{Moseley1984Thermal}%
  \BibitemOpen
  \bibfield  {author} {\bibinfo {author} {\bibfnamefont {S.~H.}\ \bibnamefont
  {Moseley}}, \bibinfo {author} {\bibfnamefont {J.~C.}\ \bibnamefont {Mather}},
  \ and\ \bibinfo {author} {\bibfnamefont {D.}~\bibnamefont {McCammon}},\
  }\href {\doibase 10.1063/1.334129} {\bibfield  {journal} {\bibinfo  {journal}
  {J. Appl. Phys.}\ }\textbf {\bibinfo {volume} {56}},\ \bibinfo {pages} {1257}
  (\bibinfo {year} {1984})}\BibitemShut {NoStop}%
\bibitem [{\citenamefont {Vijay}\ \emph {et~al.}(2011)\citenamefont {Vijay},
  \citenamefont {Slichter},\ and\ \citenamefont
  {Siddiqi}}]{Vijay2011Observation}%
  \BibitemOpen
  \bibfield  {author} {\bibinfo {author} {\bibfnamefont {R.}~\bibnamefont
  {Vijay}}, \bibinfo {author} {\bibfnamefont {D.~H.}\ \bibnamefont {Slichter}},
  \ and\ \bibinfo {author} {\bibfnamefont {I.}~\bibnamefont {Siddiqi}},\ }\href
  {\doibase 10.1103/physrevlett.106.110502} {\bibfield  {journal} {\bibinfo
  {journal} {Phys. Rev. Lett.}\ }\textbf {\bibinfo {volume} {106}},\ \bibinfo
  {pages} {110502} (\bibinfo {year} {2011})}\BibitemShut {NoStop}%
\bibitem [{\citenamefont {Andr\'{e}}\ \emph {et~al.}(1999)\citenamefont
  {Andr\'{e}}, \citenamefont {M\"{u}ck}, \citenamefont {Clarke}, \citenamefont
  {Gail},\ and\ \citenamefont {Heiden}}]{Andre1999Radiofrequency}%
  \BibitemOpen
  \bibfield  {author} {\bibinfo {author} {\bibfnamefont {M.-O.}\ \bibnamefont
  {Andr\'{e}}}, \bibinfo {author} {\bibfnamefont {M.}~\bibnamefont {M\"{u}ck}},
  \bibinfo {author} {\bibfnamefont {J.}~\bibnamefont {Clarke}}, \bibinfo
  {author} {\bibfnamefont {J.}~\bibnamefont {Gail}}, \ and\ \bibinfo {author}
  {\bibfnamefont {C.}~\bibnamefont {Heiden}},\ }\href {\doibase
  10.1063/1.124486} {\bibfield  {journal} {\bibinfo  {journal} {Appl. Phys.
  Lett.}\ }\textbf {\bibinfo {volume} {75}},\ \bibinfo {pages} {698} (\bibinfo
  {year} {1999})}\BibitemShut {NoStop}%
\bibitem [{\citenamefont {Macklin}\ \emph {et~al.}(2015)\citenamefont
  {Macklin}, \citenamefont {O'Brien}, \citenamefont {Hover}, \citenamefont
  {Schwartz}, \citenamefont {Bolkhovsky}, \citenamefont {Zhang}, \citenamefont
  {Oliver},\ and\ \citenamefont {Siddiqi}}]{Macklin2015Nearquantumlimited}%
  \BibitemOpen
  \bibfield  {author} {\bibinfo {author} {\bibfnamefont {C.}~\bibnamefont
  {Macklin}}, \bibinfo {author} {\bibfnamefont {K.}~\bibnamefont {O'Brien}},
  \bibinfo {author} {\bibfnamefont {D.}~\bibnamefont {Hover}}, \bibinfo
  {author} {\bibfnamefont {M.~E.}\ \bibnamefont {Schwartz}}, \bibinfo {author}
  {\bibfnamefont {V.}~\bibnamefont {Bolkhovsky}}, \bibinfo {author}
  {\bibfnamefont {X.}~\bibnamefont {Zhang}}, \bibinfo {author} {\bibfnamefont
  {W.~D.}\ \bibnamefont {Oliver}}, \ and\ \bibinfo {author} {\bibfnamefont
  {I.}~\bibnamefont {Siddiqi}},\ }\href {\doibase 10.1126/science.aaa8525}
  {\bibfield  {journal} {\bibinfo  {journal} {Science}\ }\textbf {\bibinfo
  {volume} {350}},\ \bibinfo {pages} {307} (\bibinfo {year}
  {2015})}\BibitemShut {NoStop}%
\end{thebibliography}%

\end{document}